\documentclass[sigconf]{acmart}

\usepackage{textcomp}
\usepackage{stfloats}
\usepackage{url}
\usepackage{verbatim}
\usepackage{amsthm}
\usepackage{subfigure}
\usepackage{makecell}
\usepackage{multirow}
\usepackage{enumitem}
\usepackage{subcaption}
\usepackage{booktabs}

\usepackage{color, colortbl}

\definecolor{Gray}{rgb}{0.9, 0.9, 0.9}

\usepackage[capitalize,noabbrev]{cleveref}

\usepackage{algorithm}
\usepackage{algorithmic}

\usepackage[inkscapelatex=false]{svg}

\AtBeginDocument{%
  }

\copyrightyear{2025}
\acmYear{2025}
\setcopyright{acmlicensed}\acmConference[SIGIR '25]{Proceedings of the 48th International ACM SIGIR Conference on Research and Development in Information Retrieval}{July 13--18, 2025}{Padua, Italy}
\acmBooktitle{Proceedings of the 48th International ACM SIGIR Conference on Research and Development in Information Retrieval (SIGIR '25), July 13--18, 2025, Padua, Italy}
\acmDOI{10.1145/3726302.3729927}
\acmISBN{979-8-4007-1592-1/2025/07}

\begin{document}

\title[COHESION]{COHESION: Composite Graph Convolutional Network with Dual-Stage Fusion for Multimodal Recommendation}

\author{Jinfeng Xu}
\affiliation{%
  {\institution{The University of Hong Kong}}
  \city{HongKong SAR}
  \country{China}}
\email{jinfeng@connect.hku.hk}

\author{Zheyu Chen}
\affiliation{%
  {\institution{The Hong Kong Polytechnic University}}
  \city{HongKong SAR}
  \country{China}}
\email{zheyu.chen@connect.polyu.hk}

\author{Wei Wang}
\affiliation{%
  {\institution{Shenzhen MSU-BIT University}}
  \city{Shenzhen}
  \country{China}}
\email{ehomewang@ieee.org}

\author{Xiping Hu}
\affiliation{%
  {\institution{Beijing Institute of Technology}}
  \city{Beijing}
  \country{China}}
\email{huxp@bit.edu.cn}

\author{Sang-Wook Kim}
\affiliation{%
  {\institution{Hanyang University}}
  \country{South Korea}}
\email{wook@hanyang.ac.kr}

\author{Edith C. H. Ngai}
\authornote{Corresponding authors}
\affiliation{%
  {\institution{The University of Hong Kong}}
  \city{HongKong SAR}
  \country{China}}
\email{chngai@eee.hku.hk}

\renewcommand{\shortauthors}{Jinfeng Xu et al.}


 
\begin{abstract}
Recent works in multimodal recommendations, which leverage diverse modal information to address data sparsity and enhance recommendation accuracy, have garnered considerable interest. Two key processes in multimodal recommendations are modality fusion and representation learning. Previous approaches in modality fusion often employ simplistic attentive or pre-defined strategies at early or late stages, failing to effectively handle irrelevant information among modalities. In representation learning, prior research has constructed heterogeneous and homogeneous graph structures encapsulating user-item, user-user, and item-item relationships to better capture user interests and item profiles. Modality fusion and representation learning were considered as two independent processes in previous work. This paper reveals that these two processes are complementary and can support each other. Specifically, powerful representation learning enhances modality fusion, while effective fusion improves representation quality. Stemming from these two processes, we introduce a \underline{CO}mposite grap\underline{H} convolutional n\underline{E}twork with dual-stage fu\underline{SION} for the multimodal recommendation, named COHESION. Specifically, it introduces a dual-stage fusion strategy to reduce the impact of irrelevant information, refining all modalities using behavior modality in the early stage and fusing their representations at the late stage. It also proposes a composite graph convolutional network that utilizes user-item, user-user, and item-item graphs to extract heterogeneous and homogeneous latent relationships within users and items. Besides, it introduces a novel adaptive optimization to ensure balanced and reasonable representations across modalities. Extensive experiments on three public datasets demonstrate the significant superiority of COHESION over various competitive baselines.

\end{abstract}

\thanks{\noindent This work was supported by the Hong Kong UGC General Research Fund no. 17203320 and 17209822, and the project grants from the HKU-SCF FinTech Academy.}



\maketitle

\section{Introduction}
\label{sec:intro}
The rapid development of the Internet has led to information explosion, making recommender systems an indispensable tool in human society. Traditional recommender systems rely on modeling user preferences through historical user-item interactions \cite{koren2021advances,xu2024improving,chen2025don, xu2024aligngroup}. However, the data sparsity problem always frustrates the accuracy of recommendations. As the types of information on social media become increasingly diverse, multimodal information has recently been used to alleviate the data sparsity problem \cite{lei2023learning,mu2022learning,xu2024mentor}. A line of work \cite{he2016vbpr,chen2017attentive} integrates multimodal information as side information to enhance item representations. To better learn the representation, many works \cite{wang2021dualgnn,wei2019mmgcn,xu2024fourierkan} construct the user-item bipartite graph and utilize the graph convolution network (GCN) to enhance the representation learning in multimodal recommendations. Despite some previous works achieving notable recommendation performance via constructing the heterogeneous user-item graph, the relationships between user-user and item-item remain unexplored. Recently, DualGNN \cite{wang2021dualgnn} constructs the user-user graph to explore the hidden preference pattern by similar users. LATTICE \cite{zhang2021mining} builds the item-item graph to capture semantically correlated signals. Both of these two homogeneous graphs improve the performance of recommendations. In this paper, we pose two comprehensive questions: \textit{\textbf{Would exploiting both user-user and item-item graphs achieve even better performance? And how?}} To answer this question, we select the most standard GCN-based multimodal recommendation model MMGCN\footnote[1]{MMGCN is the earliest and simplest graph-based multimodal recommendation model, which only contains GCNs without extra components.} \cite{wei2019mmgcn} and carefully design its three variants\footnote[2]{All of the user-user graph, item-item graph, early fusion strategy, and late fusion strategy in this investigation are designed following a recent survey paper \cite{zhou2023comprehensive}.}: MMGCN$^{u}$, MMGCN$^{i}$, and MMGCN$^{ui}$, which incorporate only user-user, only item-item, and both user-user and item-item graphs, respectively. To assess the impact of fusion strategies, we implement the early fusion and late fusion strategies for each MMGCN variant, namely MMGCN$_{e}$ and MMGCN$_{l}$, respectively. Table~\ref{tab:1} shows the recommendation performance of eight MMGCN variants on the Baby dataset. In Table~\ref{tab:1}, our findings include: 1) The late fusion strategy outperforms the early fusion strategy when integrating only user-user or item-item graphs. Conversely, the early fusion strategy is more effective than late fusion when both graph types are integrated, possibly because late fusion permits deeper modality-specific learning, but it is more susceptible to the adverse effects of irrelevant inter-modality information. 2) Among the early fusion strategies, MMGCN$^{ui}$ shows performance superior to MMGCN$^u$ and MMGCN$^i$, suggesting that combining both user and item graphs facilitates more effective representation learning. These observations suggest that suboptimal fusion strategies in prior studies may have constrained the representation learning capabilities of the otherwise well-designed models. This hypothesis is empirically supported by the results presented in Table~\ref{tab:vtm}.

To liberate the representation learning ability of composite graph convolutional network, we introduce a \underline{CO}mposite grap\underline{H} convolutional n\underline{E}twork with dual-stage fu\underline{SION} for a multimodal recommendation, named COHESION. Specifically, we propose a novel dual-stage fusion strategy to mitigate the negative effects of irrelevant information among different modalities by refining all modalities using behavior modality in the early stage and fusing all modalities' representations in the late stage. We believe that refining all modalities using behavior modality can enable each modality to map to a better semantic space. This is because the goal of multimodal recommendation is to predict user preferences instead of understanding multimodal content. Moreover, we propose the composite graph convolutional network to learn the representations for all modalities, including homogeneous and heterogeneous graphs. More specifically, we utilize a user-item graph to mine the relations between users and items and utilize user-user and item-item graphs to extract latent relations within users and items, respectively. Thanks to our dual-stage fusion strategy, our composite graph convolutional network liberates outstanding representation learning ability. Finally, we fuse all modality representations in the late stage and propose an adaptive optimization, which can make the representations learned from different modalities balanced. Extensive experiments on three widely used datasets demonstrate the superiority of COHESION over various competitive baselines. 



\begin{table}[!t]
    \centering
\caption{Performance of MMGCN variants in Baby dataset.}
\small
\label{tab:1}
\vskip -0.1in
    \begin{tabular}{c|cccc}
    \toprule
        Metrics& Recall@10 & NDCG@10 & Recall@20 & NDCG@20 \\
        \midrule
        MMGCN$_e$ & 0.0351 & 0.0589 & 0.0192 & 0.0249 \\
        MMGCN$_e^{u}$ & 0.0454 & 0.0710 & 0.0242 & 0.0311 \\
        MMGCN$_e^{i}$ & 0.0592 & 0.0924 & 0.0303 & 0.0399 \\
        MMGCN$_e^{ui}$ & 0.0630 & 0.0981 & 0.0330 & 0.0426 \\
        \midrule
        MMGCN$_l$ & 0.0378 & 0.0615 & 0.0200 & 0.0261 \\
        MMGCN$_l^{u}$ & 0.0493 & 0.0747 & 0.0266 & 0.0287 \\
        MMGCN$_l^{i}$ & 0.0627 & 0.0992 & 0.0330 & 0.0424 \\
        MMGCN$_l^{ui}$ & 0.0601 & 0.0940 & 0.0313 & 0.0401 \\
    \bottomrule
    \end{tabular}
      \vskip -0.1in
\end{table}

\section{Related work}
\subsection{Multimodal Recommendation}
Multimodal recommendation systems endeavor to harness multimodal information to enhance the accuracy of recommendations and to address the issue of data sparsity. Prior works \cite{he2016vbpr,chen2017attentive} alleviate the data sparsity challenge by extracting and utilizing visual content to enrich the representation of items based on matrix factorization techniques. Tran et al. \cite{tran2022aligning} extend this approach by incorporating textual content to further enrich item representations. More recently, several studies \cite{wang2021dualgnn,zhou2023tale} integrate information from multiple modalities concurrently. Drawing inspiration from traditional recommendation systems, Wei et al. \cite{wei2019mmgcn,wei2020graph} enhance the quality of user-item interactions through the implementation of a bipartite graph structure. To more effectively extract and analyze relationships between users and between items, DualGNN \cite{wang2021dualgnn} introduces a user-user homogeneous graph to uncover latent relationships among users. Similarly, FREEDOM \cite{zhou2023tale} and LATTICE \cite{zhang2021mining} incorporate item-item homogeneous graphs to augment modality-specific information. MONET \cite{kim2024monet} and MARIO \cite{kim2022mario} leverage modality-aware attention and tailored GCN to explore items' multimodal features. In the realm of exploring advanced structural possibilities, LGMRec \cite{guo2024lgmrec} and DiffMM \cite{jiang2024diffmm} investigate the potential of hyper-graph structures and diffusion models, respectively, to significantly enhance the effectiveness of multimodal recommendation systems. To better alleviate the data sparsity problem, numerous studies \cite{tao2022self,zhou2023bootstrap,xu2024mentor} integrate self-supervised learning tasks within the multimodal recommendation framework to discern hidden preference patterns across different modalities. However, it is important to note that self-supervised learning can also introduce irrelevant noisy information, which may include augmented self-supervised signals derived from user misclick behaviors or popularity biases, potentially reducing the model accuracy.

\subsection{Multimodal Fusion}
Multimodal information can effectively alleviate the data sparsity problem, and improve the accuracy of recommendations. However, multimodal information may lead to modal noise due to the feature of different modalities existing in different semantic spaces. Therefore, multimodal fusion plays a crucial role in noise mitigation. According to the timing of the modality fusion, most existing methods could be divided into two types: early fusion and late fusion. The early fusion methods \cite{he2016vbpr,zhang2021mining} fuse the modality features with ID embedding before constructing the heterogeneous graph. However, the modality-specific features will be lost during training. Differently, the late fusion methods \cite{wei2019mmgcn,wang2021dualgnn,zhou2023tale} learn the representation of each modality, and fuse all the predicted ratings in the end. 

With the rapid development of representation learning ability in recent multimodal recommendation models, predefined fusion strategies are more fragile to the negative impact of irrelevant information among modalities and are difficult to fuse different modalities effectively. To this end, we propose a tailored dual-stage fusion strategy, which effectively liberates the representation learning ability for multimodal recommendation models.
\section{Methodology}

\begin{figure*}[h]
    \centering
    \includegraphics[width=1\linewidth]{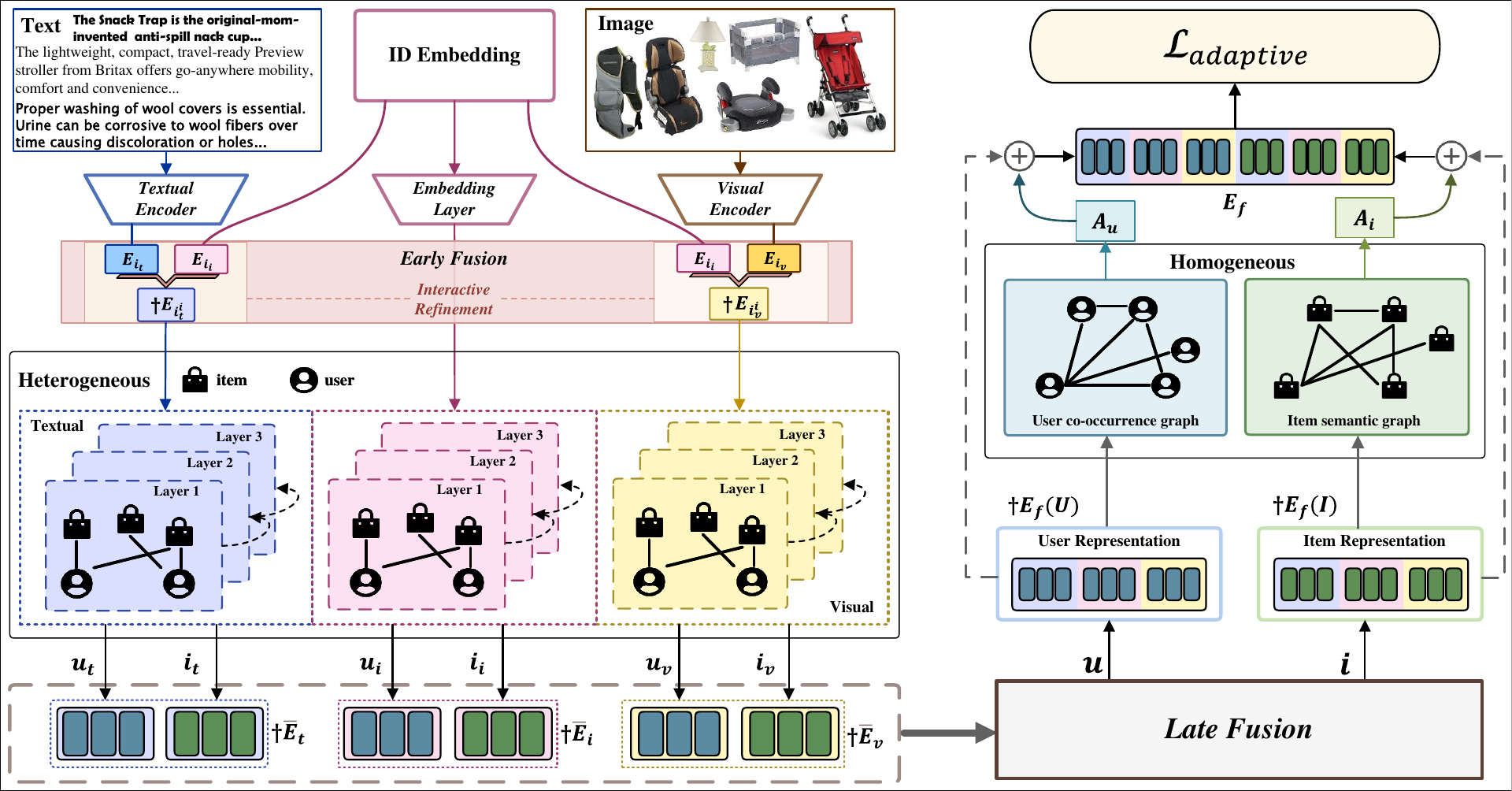}
    \vskip -0.1in
    \caption{The overview structure of COHESION.}
    \label{COHESION}
\end{figure*}

In this section, we present our proposed COHESION\footnote{The code will be released at: \href{https://github.com/Jinfeng-Xu/COHESION}{https://github.com/Jinfeng-Xu/COHESION} after the SIGIR'25 conference.} in detail. The overall architecture is shown in Fig.~\ref{COHESION}. 

\subsection{Preliminary}
Let a set of users $u \in \mathcal{U}$ and a set of items $i$ = $\{i_m|i_i, i_t, i_v\} \in \mathcal{I}$, where $m \in \mathcal{M}$ is the modality, $\mathcal{M}$ is the set of modalities, and $i_i$, $i_t$, and $i_v$ represent the behavior, textual, and visual modality features, respectively.  $\mathcal{G}$ = $(\mathcal{V}, \mathcal{E})$ be a given graph with a node set $\mathcal{V}$ and an edge set $\mathcal{E}$, where $|\mathcal{V}|$ = $|\mathcal{U}|$ + $|\mathcal{I}|$. The user-item interaction matrix is denoted as $\mathcal{R} \in \mathbb{R}^{|\mathcal{U}| \times |\mathcal{I}|}$. Specifically, each entry $\mathcal{R}_{u,i}$ indicates whether the user $u$ is connected to item $i$, with a value of 1 representing a connection and 0 otherwise. The graph structure of $\mathcal{G}$ can be denoted as the adjacency matrix $\mathcal{A} \in \mathbb{R}^{(|\mathcal{U}|+|\mathcal{I}|) \times (|\mathcal{U}|+|\mathcal{I}|)}$:
\begin{equation}
\mathcal{A}=\left[\begin{array}{cc}
0^{|\mathcal{U}| \times |\mathcal{U}|} & \mathcal{R} \\
 \mathcal{R}^T & 0^{|\mathcal{I}| \times |\mathcal{I}|}
\end{array}\right].
\end{equation}

The symmetrically normalized matrix is $\mathcal{\tilde{A}}$ $=$ $\mathcal{D}^{-\frac{1}{2}}\mathcal{A}\mathcal{D}^{-\frac{1}{2}}$, where $\mathcal{D}$ represents a diagonal degree matrix. It is worth noting that we only consider three types of modalities in this paper, including behavior, textual, and visual modalities. However, we note it can be extended to more modalities easily.

\subsection{Early Fusion}
Multimodal information can effectively alleviate the data-sparsity problem for traditional recommendations. However, it also inevitably contains irrelevant information among modalities \cite{xv2024improving,tang2019adversarial}. To this end, we propose an early fusion strategy, which refines all modalities by behavior modality before constructing heterogeneous graphs. Our early fusion strategy can effectively build a bridge to mitigate the gap between different modalities by behavior modality. In particular, we first transform the feature embeddings $E_{i_m}$ = $\{E_{i_i}, E_{i_t}, E_{i_v}\}$ into the same dimension: 
\begin{equation}
\label{eq:1}
    \dagger{E}_{i_m} = W_m^\rhd\sigma(W_m^\lhd E_{i_m} + b_1) + b_2, 
\end{equation}
where $W_m^\lhd \in \mathbb{R}^{d_m\times 4d_i}$, $W_m^\rhd \in \mathbb{R}^{4d_i\times d_i}$, $b_1$, and $b_2$ are trainable parameters. $d_m$ and $d_i$ denote embedding dimensionality of modality $m$ and behavior modality $i$, respectively. $\sigma(\cdot)$ denotes the Leaky-ReLU function \cite{xu2020reluplex}.
Then we refine each modality embedding $\dagger{E_{i_m}}$ by behavior modality $\dagger{E_{i_i}}$ respectively:
\begin{equation}
\label{eq:2}
    \dagger{E}_{i_m^i} = \sqrt{|0.5 \times ((\dagger{E}_{i_m})^2 + (\dagger{E}_{i_i})^2) + \epsilon|}, 
\end{equation}
where $\dagger{E}_{i_m^i}$ denotes refined modality embedding for modality $m$, and $\epsilon$ is a positive infinitesimal to prevent zero vector. Through this tailored yet simple early fusion strategy, we enhance the representation learning capability of multimodal recommendation models by refining different modalities via the behavior modality. This early fusion strategy mitigates the gap between modalities, leading to more accurate and robust recommendations.

\subsection{Heterogeneous Graph}
To better learn the modality-specific user and item representations, we construct a user-item graph for each modality, which is denoted as $\mathcal{G}_m$ = $\{\mathcal{G}_i, \mathcal{G}_t, \mathcal{G}_v\}$. Following previous works \cite{wei2019mmgcn,zhou2023tale}, we maintain the same graph structure $\mathcal{G}$ for different $\mathcal{G}_m$, but only retain the node features associated with a specific modality $m$. To alleviate the over-smoothing problem, we design a residual-based GCN for each user-item graph. The message propagation stage at the $l$-th graph convolution layer can be formulated as: 
\begin{equation}
\label{eq:3}
    \dagger{E}^{(l)}_m = (\operatorname{Sim}(\mathcal{\tilde{A}} {\dagger{E}^{(l-1)}_m}, \dagger{E}^{(0)}_m) + \epsilon) \times \mathcal{\tilde{A}} {\dagger{E}^{(l-1)}_m},
\end{equation}
\noindent where $\dagger{E}^{(l)}_m$ denotes the refined representations of users and items in $l$-th graph convolution layer for the modality $m$, and $\dagger{E}^{(0)}_m$ $=$ $\operatorname{Con}(E_{u_m},\dagger{E}_{i_m^i})$ is the initial layer embedding, which concatenated by random initialed user embedding $E_{u_m}$ and refined modality embedding $\dagger{E}_{i_m^i}$. $\mathcal{\tilde{A}}$ is a symmetrically normalized matrix, $\operatorname{Sim}(\cdot)$ denotes the cosine similarity function, and $\epsilon$ is a small positive bias to avoid a zero vector. The final embedding for each modality is updated by summation:
\begin{equation}
\label{eq:4}
    \dagger{\bar{E}}_m = \sum_{l=0}^L\dagger{E}^{(l)}_m,
\end{equation}
\noindent where $L$ is the total number of layers.

\subsection{Late Fusion}
In Section~\ref{sec:intro}, we explored the impact of fusion strategies on representation learning. Previous work \cite{zhou2023enhancing} also pointed out that existing fusion strategies may fail to capture modality-specific features and even corrupt the learned single-modality representation. To improve the quality of the modality fusion, we propose a tailored early fusion strategy to refine all modalities via behavior modality, which effectively alleviates the gap between modalities. After utilizing heterogeneous graphs to extract the hidden relationships between users and items, we fuse all the refined modality representations using attentive weights. Notably, our work is the first to effectively combine both early fusion and late fusion strategies to enhance the models' representation learning capabilities.
\begin{equation}
\label{eq:5}
    \dagger{E}_f = \operatorname{Con}(\alpha_m \times \dagger{\bar{E}}_m|m=i,t,v),
\end{equation}
where $\dagger{E}_f$ is the final fused representation for all refined modalities.  $\operatorname{Con}(\cdot)$ denotes the adaptive attention concatenation, and $\alpha_m$ is a learnable attention weight for each modality.

Conducting late fusion before enhancing the representations of users and items by homogeneous graphs can avoid the $|\mathcal{M}|$ times computation and storage cost of constructing homogeneous graphs for each modality separately and allow homogeneous graphs to capture features with modal synergy\footnote[4]{Partial information decomposition \cite{williams2010nonnegative,bertschinger2014quantifying} states the multivariate mutual information is decomposed into three forms of interaction (Uniqueness, Redundancy, and Synergy). Recent study \cite{dufumier2024align} finds that interaction across different modalities allows homogeneous graphs to capture features with modal synergy.} \cite{zhou2023tale,dufumier2024align}.

\subsection{Homogeneous Graph}
To improve the recommendation performance by enhancing the representations for users and items, we construct the user-user and item-item homogeneous graphs.
\subsubsection{User-User Graph}
We construct the user-user graph to explore relations between users. Considering the computation overhead, we only capture a subset of users with the most similar preferences. To this end, the pairwise cosine similarity between all users is calculated as:
\begin{equation}
\label{eq:6}
    s_{u,\tilde{u}}=\frac{(e_{u})^{\top} e_{\tilde{u}}}{\|e_{u}\|\|e_{\tilde{u}}\|}.
\end{equation}

For each user $u$, we retain the edges with top-$k$ users $\mathcal{S}_u^k$ = $\{s_{u,\check{u}}|u,\check{u} \in \mathcal{U}\}$ among all user-user pairs:
\begin{equation}
\label{eq:7}
s_{u,\check{u}} = \begin{cases}s_{u,\tilde{u}} & \text { if } s_{u,\tilde{u}} \in \text { top-} k(s_{u,v}|v \in \mathcal{U}) \\ 0 & \text { otherwise }\end{cases},
\end{equation}
where $s_{u,v}|v \in \mathcal{U}$ represents the neighbor scores
of $v$ for the user $u$. For the constructed user-user graph $\mathcal{G}^u$ = $\{\mathcal{U}, \mathcal{S}_u^k\}$, we apply the Softmax function for propagation:
\begin{equation}
\label{eq:8}
A^{(l)}_u = A^{(l-1)}_u + \sum_{s_{u,\check{u}}\in \mathcal{S}_u^k}\frac{\exp \left(s_{u, \check{u}}\right)}{\sum_{s_{u,\tilde{u}}\in \mathcal{S}_u} \exp \left(s_{u, \tilde{u}}\right)} A_{\tilde{u}}^{(l-1)},
\end{equation}
where $A^{(l)}_u$ is $l$-th layer representation of user $u$ learned from $\mathcal{G}^u$. $A^{(0)}_u$ is the user side $E_f(U)$ of $\dagger{E}_f$, which is obtained by $\dagger{E}_f[:|\mathcal{U}|]$.

\subsubsection{Item-Item Graph}
To explore relations between items, we construct an item-item graph to capture semantically correlative signals. However, the dense graph introduces huge noise, and the sparse graph leads to modality features under-exploited \cite{wang2021dualgnn}. Therefore, we only consider the top-$k$ most similar items $\mathcal{S}_i^k$ = $\{s_{i,\check{i}}|i,\check{i} \in \mathcal{I}\}$ for each item $i \in \mathcal{I}$. To this end, the pairwise cosine similarity between all items is calculated as follows:
\begin{equation}
\label{eq:9}
    s_{i,\tilde{i}}=\frac{(s_{i})^{\top} s_{\tilde{i}}}{\|s_{i}\|\|s_{\tilde{i}}\|}.
\end{equation}

Same as the user-user graph, for each item $i$, we retain the edges with top-$k$ items $\mathcal{S}_{i}^k$ = $\{s_{i,\check{i}}|i,\check{i} \in \mathcal{I}\}$ among all item-item pairs:
\begin{equation}
\label{eq:10}
s_{i,\check{i}} = \begin{cases}s_{i,\tilde{i}} & \text { if } s_{i,\tilde{i}} \in \text { top-} k(s_{i,j}|j \in \mathcal{I})\\ 0 & \text { otherwise }\end{cases},
\end{equation}
where $s_{i,j}|j \in \mathcal{I}$ represents the neighbor scores
for the item $i$. We further construct a item semantic graph $\mathcal{G}^i$ = $\{\mathcal{I}, \mathcal{S}_i^k\}$, and apply weighted aggregation function for propagation:
\begin{equation}
\label{eq:11}
A^{(l)}_i =\sum_{s_{i,\check{i}}\in S_i^k}s_{i,\check{i}} \times A_{\check{i}}^{(l-1)},
\end{equation}
where $A^{(l)}_i$ is $l$-th layer representation of item $i$ learned from $\mathcal{G}^i$. $A^{(0)}_i$ is the item side $E_f(I)$ of $\dagger{E}_f$, which can be obtained by $\dagger{E}_f[|\mathcal{U}|:]$. Then, we enhance the user and item representations by user-user and item-item graphs, respectively:
\begin{equation}
\label{eq:12}
    E_f(U) = \dagger{E}_f(U) + A_u^{L_{u}}, \quad
    E_f(I) = \dagger{E}_f(I) + A_i^{L_{i}},
\end{equation}
\begin{equation}
\label{eq:13}
    E_f = \operatorname{Con}(E_f(U)|E_f(I)),
\end{equation}
where $A_u^{L_{u}}$ and $A_i^{L_{i}}$ are the final layer results for the user-user graph and the item-item graph, respectively. $\operatorname{Con}(\cdot)$ denotes a concatenation operation. $E_f(U)$ denotes the final user representation, which can be obtained by $\dagger{E}_f[:|\mathcal{U}|]$, and $E_f(I)$ denotes the final item representation, which can be obtained by $\dagger{E}_f[|\mathcal{U}|:]$.

\subsection{Adaptive Optimization}
Bayesian Personalized Ranking (BPR) loss \cite{rendle2012bpr} are widely adopted to optimize the trainable parameters. In particular, BPR increases the gap between user $u$'s predicted ratings of the positive item $p$ and negative item $n$ for each triplet $(u, p, n) \in \mathcal{D}$, where $\mathcal{D}$ denotes the training set. Positive item $p$ is the item that has interaction with user $u$, and negative item $n$ is the item randomly sampled from items without any interaction with user $u$. 

Previous work \cite{dong2024prompt} pointed out that the BPR loss is limited to optimizing the overall user preferences and does not directly optimize the learning of modality-specific preferences, i.e., user interests specific to each modality. As a result, there could be some modality-specific interest that is not well-learned, potentially leading to unreliable final user rating predictions. To better learn the features of all modalities, we propose an adaptive BPR loss function, which pays more attention to the weak modality. The adaptive BPR loss function is defined as follows:
\begin{equation}
\label{eq:14}
w_{u, p, n}^m = 1 - \frac{\exp (y_{u, p}^m - y_{u, n}^m)}{\sum_{m \in M} \exp (y_{u, p}^m - y_{u, n}^m )},
\end{equation}
\begin{equation}
\label{eq:15}
\mathcal{L}=-\sum_{(u, p, n) \in \mathcal{D}} \log (\sigma(\sum_{m \in M}{w_{u, p, n}^m(y_{u, p}^m - y_{u, n}^m)}))+\lambda(\|\mathbf{\Theta}\|_2^2),
\end{equation}
where $w_{u, p, n}^m$ is the weight for each modality calculated by reversing the rating gap, and the rating gap is the distance between user $u$'s predicted ratings of the positive item $y_{u, p}^m$ and the negative item $y_{u, n}^m$ for each modality. $\sigma(\cdot)$ is the Sigmoid function, $\lambda$ is a hyper-parameter for regularization term, and $\mathbf{\Theta}$ denotes model parameters. 

To provide a clearer overview of our COHESION, we summarize the learning process of COHESION in Algorithm~\ref{al}.

\begin{algorithm}
\caption{Learning Process of COHESION}
\label{al}
\begin{algorithmic} [1] 
\STATE \textbf{Input:} $\mathcal{U}$, $\mathcal{I}$, $\mathcal{M}$, $\mathcal{G}$, layer number $L$ of heterogeneous graph $\mathcal{G}$, layer number $L_u$ of user-user graph $\mathcal{G}^u$, and layer number $L_i$ of item-item graph $\mathcal{G}^i$
\STATE \textbf{Output:} Optimization loss $\mathcal{L}$
\STATE Initialize $E_u$, $E_{i_m}$;
\STATE Transform item embeddings of all modalities into the same dimension $\dagger{E}_{i_m}$ $\gets$ $E_{i_m}$ with Eq.\ref{eq:1};
\STATE (Early Fusion) Refine item embeddings of each modality by behavior modality $\dagger{E}_{i_m^i}$ $\gets$ $\dagger{E}_{i_m}$ with Eq.\ref{eq:2};
\FOR{$l = 1...L$}
    \STATE Conduct message propagation within the residual-based heterogeneous graph $\dagger{E}^{(l)}_m$ $\gets$ $\dagger{E}^{(l-1)}_m $ with Eq.\ref{eq:3};
\ENDFOR
\STATE Get embedding $\dagger{\bar{E}}_m$ for each modality with Eq.\ref{eq:4};
\STATE (Late Fusion) Attentively fuse all modality embeddings $\dagger{E}_f$ $\gets$ $\dagger{\bar{E}}_m$ with Eq.\ref{eq:5};
\STATE Construct the user-user graph $\mathcal{G}^u$ with Eq.\ref{eq:6} and Eq.\ref{eq:7}.
\STATE Construct the item-item graph $\mathcal{G}^i$ with Eq.\ref{eq:9} and Eq.\ref{eq:10}.
\FOR{$l = 1...L_{u}$}
    \STATE Conduct message propagation within the user-user homogeneous graph $A^{(l)}_u$ $\gets$ $A^{(l-1)}_u$ with Eq.\ref{eq:8};
\ENDFOR
\FOR{$l = 1...L_{i}$}
    \STATE Conduct message propagation within the item-item homogeneous graph $A^{(l)}_i$ $\gets$ $A^{(l-1)}_i$ with Eq.\ref{eq:11};
\ENDFOR
\STATE Get final enhanced embedding $E_f$ by both user-user and item-item graphs with Eq.\ref{eq:12} and Eq.\ref{eq:13}.
\STATE Calculate adaptive BPR loss $\mathcal{L}$ with Eq.\ref{eq:14} and Eq.\ref{eq:15}.
\end{algorithmic}
\end{algorithm}
\section{Experiments}

\label{sec:experiments}
In this section, we conduct extensive experiments on some widely used real-world datasets. Experiment results can evaluate the following questions: R\textbf{Q1:} How does COHESION perform compared with various state-of-the-art models? \textbf{RQ2:} How do the key components in COHESION influence the performance of recommendation accuracy? \textbf{RQ3:} Can COHESION effectively mitigate the irrelevant information among modalities? \textbf{RQ4:} How does COHESION perform compared with various state-of-the-art models in different sparse data scenarios? \textbf{RQ5:} How efficient is COHESION compared with various state-of-the-art models? \textbf{RQ6:} How sensitive is COHESION under the perturbation of hyper-parameters?


\begin{table}[h]
    \centering
\caption{Statistics of three experimented datasets with modalities item Visual(V) and Textual(T).}
\small
\vskip -0.1in
\label{tab:dataset_statistics}
    \begin{tabular}{cccccccccc}
    \hline
         Dataset&& \multicolumn{2}{c}{Baby} && \multicolumn{2}{c}{Sports} && \multicolumn{2}{c}{Clothing}\\
         \hline \hline
         Modality && V & T && V & T && V & T\\ 
         Embed Dim && 4,096 & 384 && 4,096 & 384 && 4,096 & 384\\ \cline{1-1} \cline{3-4} \cline{6-7} \cline{9-10}
         User && \multicolumn{2}{c}{19,445} && \multicolumn{2}{c}{35,598} && \multicolumn{2}{c}{39,387} \\
         Item && \multicolumn{2}{c}{7,050} && \multicolumn{2}{c}{18,357} && \multicolumn{2}{c}{23,033} \\
         Interaction && \multicolumn{2}{c}{160,792} && \multicolumn{2}{c}{296,337} && \multicolumn{2}{c}{278,677} \\ \hline
         Sparsity && \multicolumn{2}{c}{99.88\%} && \multicolumn{2}{c}{99.95\%} && \multicolumn{2}{c}{99.97\%} \\
         \hline
    \end{tabular}
\end{table}

\subsection{Settings}

\subsubsection{Datasets}
Our experiments are conducted on three real-world datasets: Baby, Sports, and Clothing from the Amazon dataset \cite{mcauley2015image}, which include visual and textual modalities for each item. Table~\ref{tab:dataset_statistics} shows the statistics of these datasets. Following most previous works \cite{yi2021multi,zhou2023bootstrap, xu2025survey}, we adopt the 5-core setting to filter users and items for each dataset. We use the pre-trained sentence-transformer to extract textual features with 384 dimensions and use the published 4096-dimensional visual features as in MMRec \cite{zhou2023mmrecsm}. For each dataset, we use the ratio 8:1:1 to split the historical interactions randomly for training, validation, and testing.

\begin{table*}[!t]
\caption{Performance comparison of baselines and COHESION in terms of Recall@K (R@K) and NDCG@K (N@K). The superscript $^*$ indicates the improvement is statistically significant where the p-value is less than 0.01.}
\vskip -0.1in
\small
\centering
\label{tab:comparison results}
    \begin{tabular}{ccccc|cccc|cccc}
    \toprule
         Datasets&  \multicolumn{4}{c}{Baby}&  \multicolumn{4}{c}{Sports}&  \multicolumn{4}{c}{Clothing}\\ \midrule
         Model& R@10& R@20& N@10& N@20 & R@10& R@20& N@10& N@20
         & R@10& R@20& N@10& N@20\\\midrule
         MF-BPR & 0.0357& 0.0575& 0.0192& 0.0249& 0.0432& 0.0653& 0.0241& 0.0298& 0.0187& 0.0279& 0.0103& 0.0126\\
         LightGCN & 0.0479& 0.0754& 0.0257& 0.0328& 0.0569& 0.0864& 0.0311& 0.0387& 0.0340& 0.0526& 0.0188& 0.0236\\
         SimGCL & 0.0513 & 0.0804 & 0.0273 & 0.0350& 0.0601 & 0.0919 & 0.0327 & 0.0414 & 0.0356 & 0.0549 & 0.0195 & 0.0244\\
         LayerGCN & 0.0529& 0.0820& 0.0281& 0.0355& 0.0594& 0.0916& 0.0323& 0.0406& 0.0371& 0.0566& 0.0200& 0.0247\\
         \midrule
         VBPR &  0.0423&  0.0663&  0.0223&  0.0284 &  0.0558&   0.0856& 0.0307& 0.0384 &  0.0281&   0.0415& 0.0158& 0.0192\\
         MMGCN &  0.0378&  0.0615&  0.0200&  0.0261 &  0.0370&   0.0605& 0.0193& 0.0254& 0.0218&   0.0345& 0.0110& 0.0142\\
         DualGNN &  0.0448&  0.0716&  0.0240&  0.0309 &  0.0568&  0.0859& 0.0310& 0.0385 &  0.0454&  0.0683& 0.0241& 0.0299\\
         SLMRec &  0.0529&  0.0775&  0.0290&  0.0353&  0.0663&  0.0990& 0.0365& 0.0450&  0.0452&  0.0675& 0.0247& 0.0303\\
         LATTICE &  0.0547&  0.0850&  0.0292&  0.0370&  0.0620&  0.0953&  0.0335&  0.0421 &  0.0492&  0.0733& 0.0268& 0.0330\\
         FREEDOM &0.0627&\underline{0.0992}&0.0330&0.0424 &0.0717&\underline{0.1089}&0.0385&\underline{0.0481} &  \underline{0.0628}&  \underline{0.0941}& \underline{0.0341}& \underline{0.0420}\\
         BM3 &0.0564&0.0883&0.0301&0.0383&0.0656&0.0980&0.0355&0.0438&  0.0422&  0.0621& 0.0231& 0.0281\\
         LGMRec &\underline{0.0639}&0.0989&\underline{0.0337}&\underline{0.0430}&\underline{0.0719}&0.1068&\underline{0.0387}& 0.0477 & 0.0555&  0.0828& 0.0302& 0.0371\\ \midrule
         COHESION &  \textbf{0.0680$^*$}&  \textbf{0.1052$^*$}& \textbf{0.0354$^*$}&  \textbf{0.0454$^*$}&  \textbf{0.0752$^*$}&  \textbf{0.1137$^*$} &\textbf{0.0409$^*$} & \textbf{0.0503$^*$} &  \textbf{0.0665$^*$}&  \textbf{0.0983$^*$} &\textbf{0.0358$^*$} & \textbf{0.0438$^*$} \\
         Improv.&  6.42\%&  6.05\%&  5.04\%&  5.58\%&  4.59\%&  4.41\%&  5.68\%&  4.57\%&  5.89\%&  4.46\%&  4.99\%&  4.28\%\\ \bottomrule
    \end{tabular}
\end{table*}

\subsubsection{Baseline}
To evaluate the effectiveness of our COHESION, we compare it with the following methods, which can be divided into two groups. We detailed all of them as follows:

\noindent1) Traditional recommendation models:
\begin{itemize}[leftmargin=*]
    \item \textbf{MF-BPR} \cite{rendle2012bpr} utilizes the BPR loss to optimize the classic collaborative filtering method, which learns user and item representations by the matrix factorization method.
    \item \textbf{LightGCN} \cite{he2020lightgcn} excludes some unnecessary components from GCN-based collaborative filtering to make it more appropriate for recommendation.
    \item \textbf{SimGCL} \cite{yu2022graph} proposes a graph contrastive learning method, which directly injects random noises into the representation.
    \item \textbf{LayerGCN} \cite{zhou2023layer} utilizes the residual connection to build a layer-refined GCN to alleviate LightGCN's over-smoothing problem.
\end{itemize}

\noindent2) Multimodal recommendation models:
\begin{itemize}[leftmargin=*]
    \item \textbf{VBPR} \cite{he2016vbpr} integrates the visual and textual features with ID embeddings as side information for each item, which can be seen as multimodal matrix factorization.
    \item \textbf{MMGCN} \cite{wei2019mmgcn} utilizes GCN for each modality to learn the modality-specific features, and then fuses all the user predicted ratings for all modalities to get the final predicted rating.
    \item \textbf{DualGNN} \cite{wang2021dualgnn} proposes a user-user graph to explore the hidden preference pattern. 
    \item \textbf{LATTICE} \cite{zhang2021mining} proposes an item-item graph to capture semantically correlated signals among items.
    \item \textbf{FREEDOM} \cite{zhou2023tale} optimizes LATTICE by freezing the item-item graph and denoising the user-item graph.
    \item \textbf{SLMRec} \cite{tao2022self} proposes a self-supervised framework for the multimodal recommendation, which builds a node self-discrimination task to uncover the hidden multimodal patterns of items.
    \item \textbf{BM3} \cite{zhou2023bootstrap} simplifies the self-supervised framework by directly perturbing the representation through a dropout mechanism.
    \item \textbf{LGMRec} \cite{guo2024lgmrec} utilizes local embeddings with local topological information and global embeddings with hypergraphs.
\end{itemize}

\subsubsection{Evaluation Metrics}
For a fair comparison, we follow the settings of previous works \cite{wang2021dualgnn,zhou2023bootstrap,zhou2023tale} to adopt two widely-used evaluation metrics for top-$K$ recommendation: Recall@K and NDCG@K. We report the average scores for all users in the test dataset under both K = 10 and K = 20, respectively.

\subsubsection{Implementation Details}
Following the existing work \cite{zhou2023mmrecsm}, we fix the user and item embedding size to 64 for all models, and then initialize them with the Xavier method \cite{glorot2010understanding}. Meanwhile, we optimize all models with Adam optimizer \cite{kingma2014adam}. For each baselines, we perform a grid search for all the hyper-parameters following its published paper to find the optimal setting. All models are implemented by PyTorch and evaluated on an RTX 4070Ti GPU card. For our proposed COHESION, we perform a grid search on the learning rate in \{1e-1, 1e-2, 1e-3, 1e-4\}, the regularization weight $\lambda$ in \{1e-1, 1e-2, 1e-3, 1e-4\}, and residual-based GCN layers $L$ in \{1, 2, 3, 4\}. We empirically fix the number of GCN layers in the homogeneous graph (both user-user graph and item-item graph) with $L_{u}/L_{i}$ = 1 \cite{zhang2021mining,wang2021dualgnn}. The $k$ of top-$k$ in the user-user graph is also empirically set as 10 \cite{zhou2023tale,zhang2021mining,wang2021dualgnn}. To avoid over-fitting, the early stopping and total epochs are fixed at 20 and 1000, respectively. Following MMRec \cite{zhou2023mmrecsm}, we use Recall@20 on the validation dataset as the indicator for best record updating.

\subsection{Performance Comparison (RQ1)}
Table~\ref{tab:comparison results} shows the performance comparison of our proposed COHESION with other baselines on three widely used datasets. The optimal and optimal results are indicated in \textbf{bold} and \underline{underlined}, respectively. We have the following observations:
\begin{itemize}[leftmargin=*]
    \item Most multimodal models achieve better performance than traditional recommendation models, which verifies that multimodal information can effectively alleviate the data sparsity problem in traditional recommendation systems.
    \item Some multimodal recommendation methods result in inconsistent performance across various scenarios. For example, LATTICE outperforms SLMRec on the Baby dataset, but is less effective on the Sports dataset. The self-supervised learning works better in the Sports dataset than in the Baby dataset.
    \item COHESION makes a huge improvement over all multimodal recommendation models. This validates the effectiveness of our COHESION. Our dual-stage fusion strategy effectively bridges different modalities by refining each modality via behavior modality. Our well-designed composite graph convolutional network leverages both heterogeneous and homogeneous graphs to improve the learning representation ability. Moreover, we introduce an adaptive optimization, which is designed to enhance the comprehensive learning of under-learned modalities. As a result, COHESION outperforms existing multimodal methods. For example, COHESION improves the strongest baseline (FREEDOM or LGMRec) in terms of R@20 on the Baby and Sports datasets by 6.05\% and 4.41\%, respectively. 
    \item DualGNN and LATTICE both achieve competitive performance by adding a user-user and item-item graph, respectively. Our COHESION significantly outperforms these two methods. We owe our superiority to our novel dual-stage fusion strategy liberates the representational capabilities of a well-designed composite graph convolutional network. 
\end{itemize}

\subsection{Ablation Study (RQ2)}

In this section, we empirically evaluate the effectiveness of different components of COHESION.
\subsubsection{Effect of Early Fusion}
To investigate the effects of the early fusion component in COHESION, we set up the following model variants:
\begin{itemize}[leftmargin=*]
    \item \textbf{COHESION-itv}: Removing the entire early fusion process.
    \item \textbf{COHESION-tv}: Removing refinement operation of textual and visual modalities in the early fusion process.
    \item \textbf{COHESION-iv}: Removing the refinement operation of behavior and visual modalities in the early fusion process.
    \item \textbf{COHESION-it}: Removing the refinement operation of behavior and textual modalities in the early fusion process.
    \item \textbf{COHESION-v}: Removing the refinement operation of visual modality in the early fusion process.
    \item \textbf{COHESION-t}: Removing the refinement operation of textual modality in the early fusion process.
    \item \textbf{COHESION-i}: Removing the refinement operation of behavior modality in the early fusion process.
\end{itemize}

\begin{figure}[!h]
\vskip -0.15in
    \centering
    \includegraphics[width=1\linewidth]{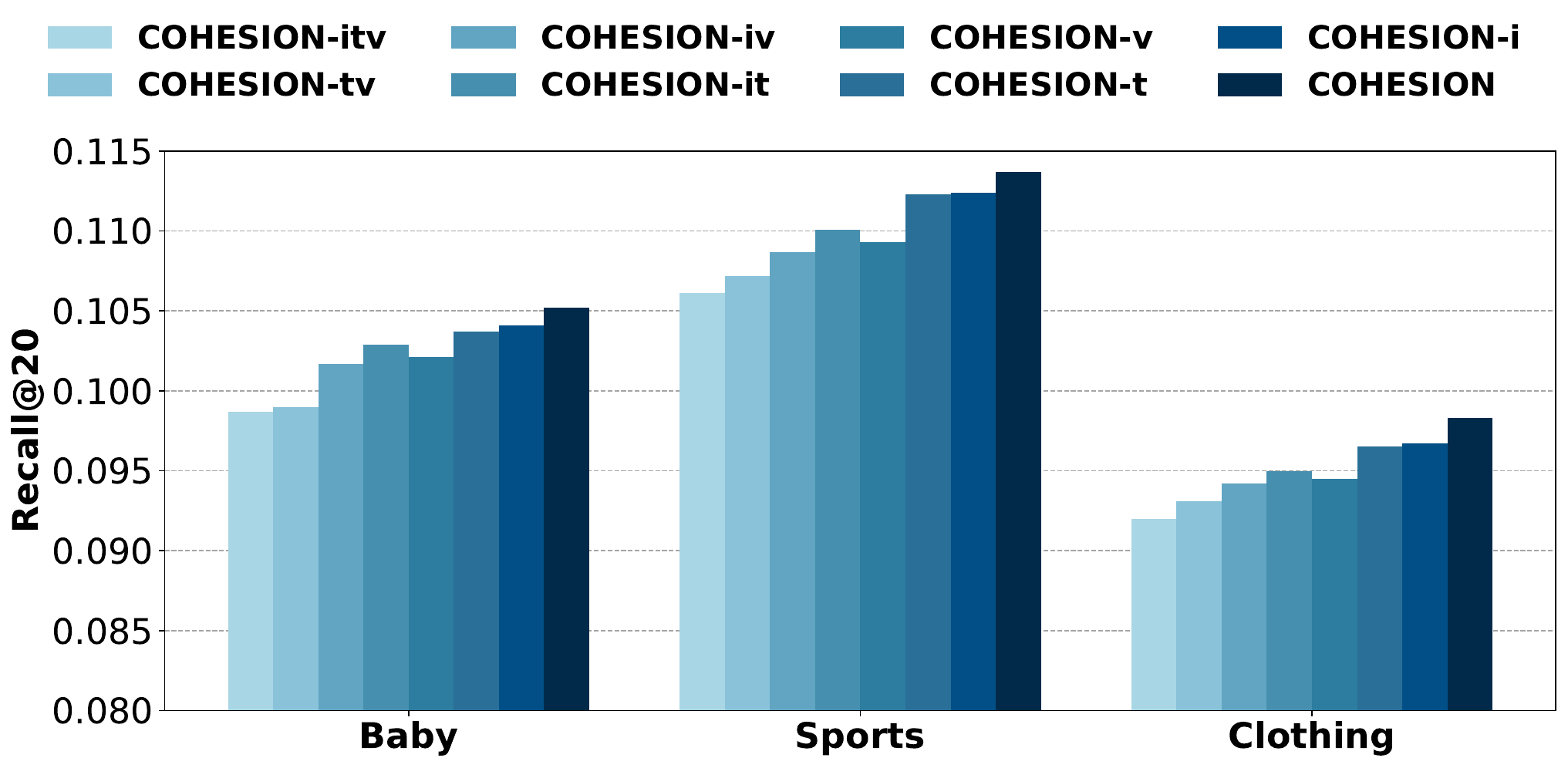}
     \vskip -0.15in
    \caption{Effect of early fusion.}
    \label{fig:interactive refinement}
    \vskip -0.15in
\end{figure}

Fig.~\ref{fig:interactive refinement} shows the results (Recall@20) of these variants and COHESION on all three datasets. We have the following observations:
\begin{itemize}[leftmargin=*]
\item  1) Results of \textbf{COHESION-itv} $<$ \textbf{COHESION-tv}, \textbf{COHESION-iv} $<$ \textbf{COHESION-v}, and \textbf{COHESION-it} $<$ \textbf{COHESION-t} showing that refining behavior modality also leads to a slight improvement in recommendation performance.

\item 2) COHESION performs better than \textbf{COHESION-i}, \textbf{COHESION-v}, and \textbf{COHESION-t}, which shows that refining any modality can lead to improve the recommendation accuracy.

\item 3) COHESION outperforms all seven variants, verifying the effectiveness of our early fusion strategy.
\end{itemize}



\begin{table}[!h]
    \centering
    \caption{Effect of homogeneous graphs in terms of Recall@K (R@K) and NDCG@K (N@K).}
    \small
    \vskip -0.1in
    \label{tab:homogeneous}
    \tabcolsep=0.06in
    \begin{tabular}{c|c|cccc}
    \toprule
        Dataset &  Variants & R@10& R@20& N@10& N@20 \\ \midrule
        \multirow{4}{*}{Baby} 
        &COHESION&    \textbf{0.0680}&\textbf{0.1052}&\textbf{0.0354}&\textbf{0.0454}\\
        &COHESION-u&  0.0660&0.1024&0.0349&0.0443\\
        &COHESION-i&  0.0564&0.0884&0.0307&0.0389\\
        &COHESION-ui& 0.0547&0.0870&0.0291&0.0375\\ \midrule
        \multirow{4}{*}{Sports} 
        &COHESION&    \textbf{0.0752}&\textbf{0.1137}&\textbf{0.0409}&\textbf{0.0503}\\
        &COHESION-u&  0.0705&0.1086&0.0384&0.0482\\
        &COHESION-i&  0.0622&0.0963&0.0337&0.0425\\
        &COHESION-ui& 0.0603&0.0937&0.0327&0.0413\\ \midrule
        \multirow{4}{*}{Clothing} 
        &COHESION&    \textbf{0.0665}&\textbf{0.0983}&\textbf{0.0358}&\textbf{0.0438}\\
        &COHESION-u&  0.0630&0.0942&0.0339&0.0417\\
        &COHESION-i&  0.0499&0.0744&0.0271&0.0331\\
        &COHESION-ui& 0.0478&0.0701&0.0262&0.0319\\ 
    \bottomrule
    \end{tabular}
    \vskip -0.1in
\end{table}

\subsubsection{Effect of Homogeneous Graphs}
To identify the effectiveness of homogeneous graphs in COHESION, we set up the following variants: 1) \textbf{COHESION-ui}, which removes both user-user and item-item graphs. 2) \textbf{COHESION-i}, which removes only the item-item graph. 3) \textbf{COHESION-u}, which removes only the user-user graph. As illustrated in Table~\ref{tab:homogeneous}, we observe that both user-user and item-item graphs could improve the performance of recommendations. The item-item graph can better represent the user's preference than the user-user graph, which is identified by the results \textbf{COHESION-ui} $<$ \textbf{COHESION-i} $<$ \textbf{COHESION-u} $<$ \textbf{COHESION}. Note that user-user and item-item graphs can improve recommendation accuracy together to achieve higher performance.

\begin{table}[!h]
    \centering
    \small
    \caption{Effect of adaptive optimization in terms of Recall@K (R@K) and NDCG@K (N@K).}
    \label{tab:adaptive optimization}
     \vskip -0.1in
    \begin{tabular}{c|l|cccc}
    \toprule
        Dataset &   Variants& R@10 & R@20 & N@10 & N@20 \\ \midrule
        \multirow{4}{*}{Baby} 
        &DualGNN&  0.0448&  0.0716&  0.0240&  0.0309\\
        &DualGNN+&  \textbf{0.0492}&  \textbf{0.0753}&  \textbf{0.0261}&  \textbf{0.0328}\\ \cmidrule{2-6}
        &LATTICE&  0.0547&  0.0850&  0.0292&  0.0370\\
        &LATTICE+&  \textbf{0.0573}&  \textbf{0.0881}&  \textbf{0.0303}&  \textbf{0.0388}\\ \midrule
        \multirow{4}{*}{Sports} 
        &DualGNN&  0.0568&  0.0859&  0.0310&  0.0385\\
        &DualGNN+&  \textbf{0.0587}&  \textbf{0.0894}&  \textbf{0.0316}&  \textbf{0.0394}\\ \cmidrule{2-6}
        &LATTICE&  0.0620&  0.0953&  0.0335&  0.0421\\
        &LATTICE+& \textbf{0.0634}& \textbf{0.0974}& \textbf{0.0343}& \textbf{0.0430}\\ \midrule
         \multirow{4}{*}{Clothing} 
        &DualGNN&  0.0454&  0.0683&  0.0241&  0.0299\\
        &DualGNN+&  \textbf{0.0499}&  \textbf{0.0737}&  \textbf{0.0270}&  \textbf{0.0334}\\ \cmidrule{2-6}
        &LATTICE&  0.0492&  0.0733&  0.0268&  0.0330\\
        &LATTICE+& \textbf{0.0524}& \textbf{0.0787}& \textbf{0.0285}& \textbf{0.0354}\\ 
        \bottomrule
    \end{tabular}
    \vskip -0.1in
\end{table}

\subsubsection{Effect of Adaptive Optimization}
Our adaptive optimization could serve as a play-and-plug component for other models to improve their performance. To evaluate the effectiveness of adaptive optimization in other models, we design the corresponding model variants, named \textbf{DualGNN+} and \textbf{LATTICE+}, where symbol \textbf{+} means using adaptive optimization. As shown in Table~\ref{tab:adaptive optimization}, our adaptive optimization achieves significant improvement for all models on all datasets.

\begin{table*}[!t]
    \centering
\caption{Performance comparison on all three datasets (Recall@20). The notation Visual means only adding visual modality, Textual means only adding textual modality, and Multimodal means adding both visual and textual modalities. Improv. denotes the improvement of adding multimodal compared to adding the optimal single modality.}
\vskip -0.1in
\label{tab:vtm}
\small
    \begin{tabular}{c|c|ccccccccc}
    \toprule
         Dataset& Modality & VBPR & MMGCN & DualGNN & SLMRec & LATTICE & FREEDOM & BM3 & LGMRec & COHESION\\ \midrule
         \multirow{4}{*}{Baby} & Visual & \textbf{0.0665} & \textbf{0.0634} & 0.0830 & 0.0789 & 0.0779 & 0.0816 & 0.0806 & 0.0954 & 0.0878 \\
         ~ & Textual & 0.0638 & 0.0623 & \textbf{0.0943} & \textbf{0.0797} & \textbf{0.0880} & 0.0980 & \textbf{0.0842} & \textbf{0.0997} & 0.1002 \\\cmidrule{2-11}
         ~ & Multimodal & 0.0663 & 0.0615 & 0.0716 & 0.0775 & 0.0848 & \textbf{0.0992} & 0.0833 & 0.0989 & \textbf{0.1052} \\
         ~ & Improv. & -0.30\% & -3.09\% & -31.17\% & -2.84\% & -3.77\% & +1.22\% & -1.08\% & -0.80\% & +4.99\%\\\midrule
          \multirow{4}{*}{Sports} & Visual & \textbf{0.0849} & 0.0574 & 0.0896 & 0.0928 & 0.0902 & 0.0879 & 0.0973 & 0.1059 & 0.1085 \\
         ~ & Textual & 0.0714 & \textbf{0.0623} & \textbf{0.0958} & \textbf{0.1011} & \textbf{0.0978} & 0.1083 & \textbf{0.0989} & \textbf{0.1093} & 0.1099 \\\cmidrule{2-11}
         ~ & Multimodal & 0.0856 & 0.0605 & 0.0859 & 0.0990 & 0.0953 & \textbf{0.1089} & 0.0980 & 0.1068 & \textbf{0.1137} \\
         ~ & Improv. & -0.82\% & -2.89\% & -10.33\% & -2.08\% & -2.56\% & +0.55\% & -0.91\% & -2.29\% & +3.46\%\\\midrule
         \multirow{4}{*}{Clothing} & Visual & \textbf{0.0433} & \textbf{0.0363} & 0.0730 & 0.0662 & 0.0705 & 0.0816 & 0.0608 & 0.0805 & 0.0928 \\
         ~ & Textual & 0.0397 & 0.0358 & \textbf{0.0784} & \textbf{0.0698} & \textbf{0.0767} & 0.0937 & \textbf{0.0650} & \textbf{0.0847} & 0.0956 \\\cmidrule{2-11}
         ~ & Multimodal & 0.0415 & 0.0345 & 0.0683 & 0.0675 & 0.0733 & \textbf{0.0941} & 0.0621 & 0.0828 & \textbf{0.0983} \\
         ~ & Improv. & -4.16\% & -4.96\% & -12.77\% & -3.30\% & -4.43\% & +0.43\% & -4.46\% & -2.24\% & +2.82\%\\

    \bottomrule
    \end{tabular}
\end{table*}

\subsection{Effectiveness of Mitigating Irrelevant Information among Modalities (RQ3)}
The negative effect of irrelevant information among modalities will decrease modal representation learning ability and even lead to a weaker performance for incorporating multiple modalities than incorporating a single modality. Table~\ref{tab:vtm} shows that most existing models suffer from this dilemma, we observe that only FREEDOM and our COHESION achieve higher performance for incorporating multiple modalities and that COHESION made a much more significant improvement of 4.99\%. It verifies that our early fusion strategy bridges different modalities to mitigate the negative effect of irrelevant information among modalities.

\begin{figure}[!t]
    \centering
    \label{fig:without dual-stage fusion}
    \includegraphics[width=1\linewidth]{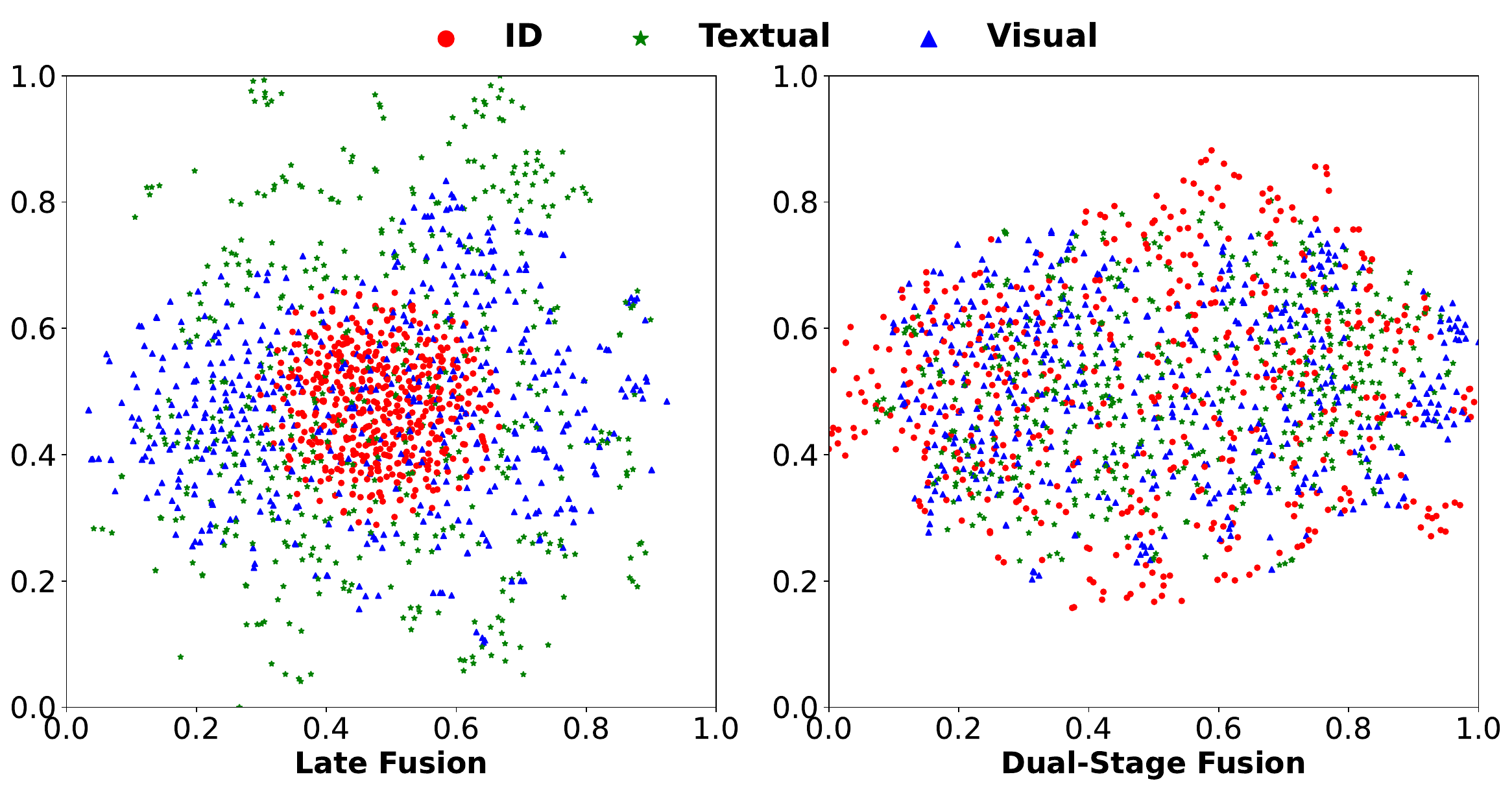}
    \vskip -0.15in
    \caption{Visualization of distribution through t-SNE. We randomly sample 500 items from Baby dataset. Red, green, and blue represent the ID, textual, and visual modalities.}
    \label{fig:t-SNE}   
     \vskip -0.2in
\end{figure}

To gain a better understanding of the advantages of our dual-stage strategy for mitigating irrelevant information among modalities, we take a step: randomly pick 500 items from the Baby dataset and use the t-SNE \cite{van2008visualizing} algorithm to map the representations of late fusion strategy and our dual-stage strategy into a two-dimensional space, respectively. Fig.~\ref{fig:t-SNE} visualizes the distributions of all modalities. The distributions of all modalities from our dual-stage fusion strategy exhibit significantly greater similarity compared to those from the late fusion strategy, further validating the effectiveness of our dual-stage fusion approach in aligning modality representations.

\begin{figure}[!t]
    \centering
    \includegraphics[width=1\linewidth]{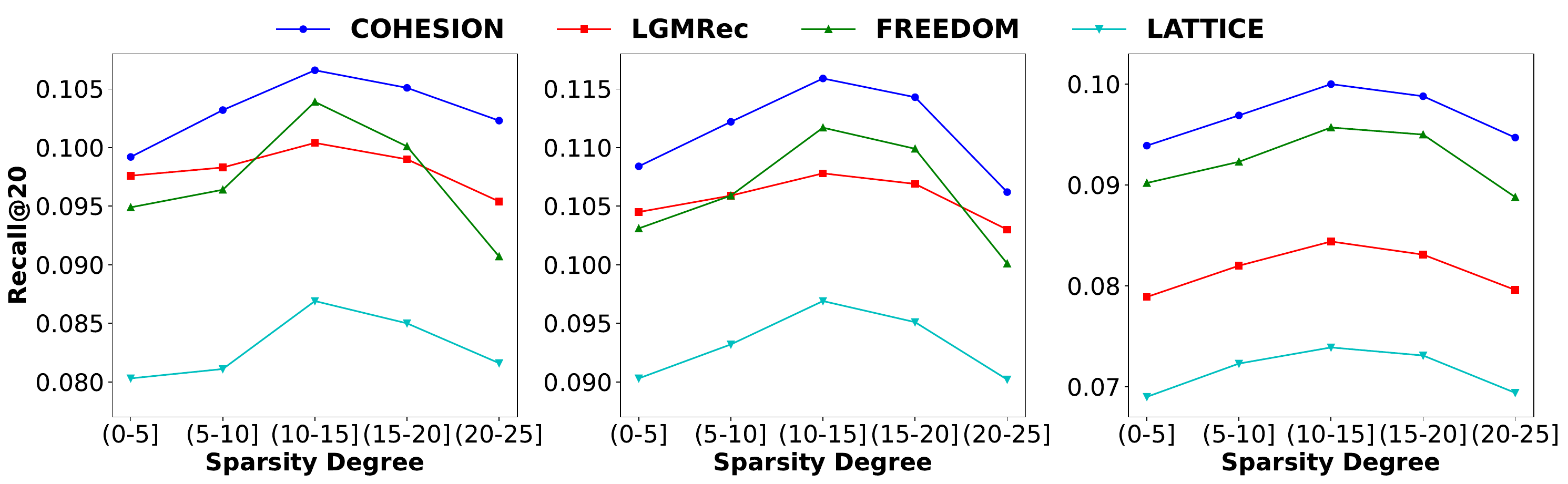}
    \vskip -0.1in
    \caption{Sparsity degree analysis on three datasets.}
    \label{fig:sparsity degree}   
     \vskip -0.15in
\end{figure}

\subsection{Different Sparse Data Scenarios (RQ4)}
In this section, we verify the effectiveness of COHESION in different sparse data scenarios. We conduct experiments on sub-datasets with varying levels of data sparsity using all three datasets. We compare the performance of COHESION against three outstanding baselines (LATTICE, FREEDOM, LGMRec). We split user groups by the number of interactions in the training set (e.g., the first group consists of users with 0-5 interacted items). Fig.~\ref{fig:sparsity degree} shows that COHESION consistently outperforms all baselines on all datasets with different degrees of sparsity, which demonstrates the effectiveness in different sparse data scenarios.

\begin{table*}[!t]
    \centering
    \small
\caption{Comparison of COHESION against competitive baselines on efficiency. Parameter denotes the learnable parameter for the model, and Time denotes the total duration of each iteration, including both the training and evaluation phases.}
\label{tab: TP}
\vskip -0.1in
    \begin{tabular}{c|l|ccccccccc}
    \toprule
        Dataset& Metrics & VBPR & MMGCN & DualGNN & SLMRec & LATTICE & FREEDOM & BM3 & LGMRec & COHESION \\
        \midrule
        \multirow{2}{*}{Baby} 
        & Time (s/epoch)& 2.12& 3.96& 5.48& 3.75& 4.13& 2.77& 2.55& 5.93& 4.47\\
        & Parameter (M)& 3.2& 1.4& 3.8& 2.0& 33.6& 33.6& 33.6& 38.6& 37.0\\
        \midrule
        \multirow{2}{*}{Sports} 
        & Time (s/epoch)& 3.55& 11.01& 15.90& 6.14& 11.90& 5.86& 5.39& 8.98& 7.91\\
        & Parameter (M)& 6.0& 1.4& 5.9& 3.8& 86.0& 86.0& 86.0& 99.9& 91.5\\ 
        \midrule
        \multirow{2}{*}{Clothing} 
        & Time (s/epoch)& 3.90& 13.04& 16.93& 6.94& 14.10& 6.29& 6.09& 10.02& 9.05\\
        & Parameter (M)& 6.8& 1.4& 6.4& 4.3& 107.5& 107.5& 107.5& 126.1& 113.4\\
    \bottomrule
    \end{tabular}
\end{table*}

\subsection{Efficiency Analysis (RQ5)}
\subsubsection{Time Complexity}
We provide a theoretical time complexity analysis for our COHESION. Firstly, the heterogeneous graph cost is $O(2$$\times$$L$$\times$$|\mathcal{G}|$$\times$$d/B)$, where $|\mathcal{G}|$ is the number of edges, $d$ is the embedding dimension, $B$ is the batch size, and $L$ is the GCN layers number. Secondly, the homogeneous graph cost is $O((L_{u}|\mathcal{U}|^2 + L_{i}|\mathcal{I}|^2)$$\times$$d)$, where $L_{u}$ and $L_{i}$ are the number of a user-user graph and an item-item graph, respectively. Lastly, the BPR loss cost is $O(2$$\times$$d$$\times$$B)$. Note that COHESION doesn't adopt self-supervised tasks, which makes it more efficient than SSL-based models. Since we conduct our late fusion strategy before enhancing representation by the user-user and item-item graphs, we save $|\mathcal{M}|$ times of computation and storage resources compared to DualGNN, LATTICE, and FREEDOM. This allows us to maintain similar computation and storage resource requirements as competitive baselines when using both the user-user and item-item graphs.

\subsubsection{Computational Resource}
Moreover, we report the training time (including both computation time and evaluating time) and model parameters for our COHESION and baselines in Table~\ref{tab: TP}. We observe that our COHESION achieves a significant improvement over competitive baselines (LATTICE, FREEDOM, BM3, and LGMRec) while requiring comparable computation resources and training time. This observation underscores the efficiency of our COHESION model, highlighting its ability to achieve superior recommendation performance without necessitating an increase in resource consumption or training time, thereby demonstrating its practical applicability in multimodal recommendation scenarios.

\subsubsection{Convergency}
\begin{figure}[!h]
    \centering
    \includegraphics[width=1\linewidth]{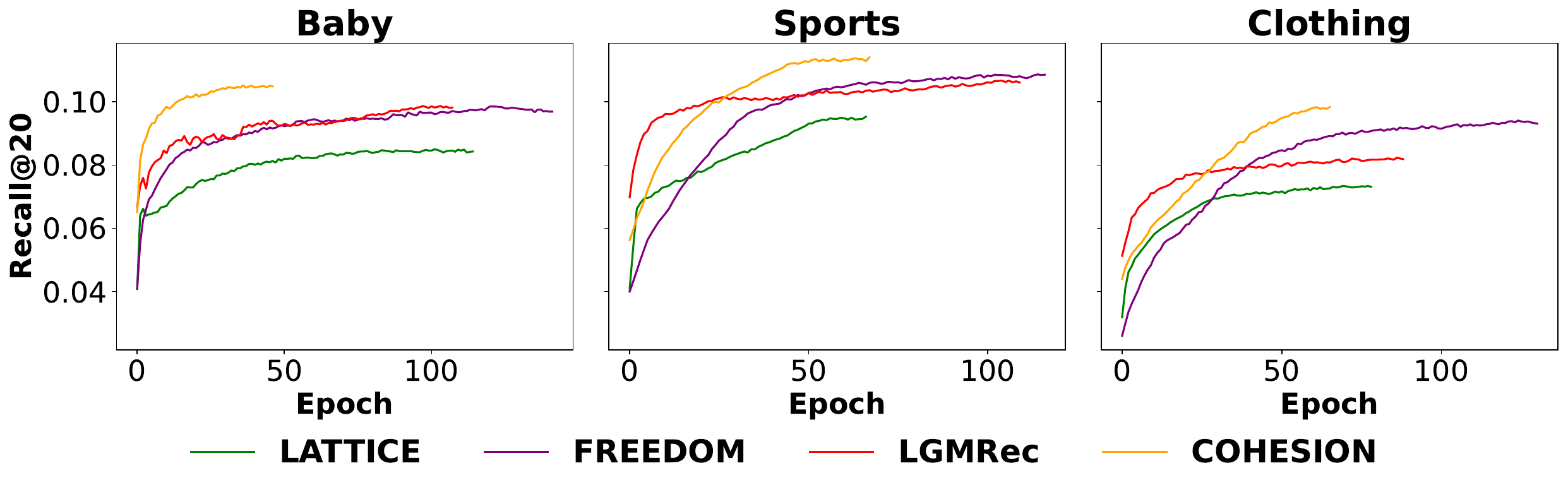}
    \vskip -0.2in
    \caption{Convergence study in terms of Recall@20.}
    \label{fig:convergency}
     \vskip -0.1in
\end{figure}

In Figure~\ref{fig:convergency}, we show the training curves of our COHESION and competitive baselines (LATTICE, FREEDOM, and LGMRec) on all three datasets, as the numbers of iterations and epochs increase. The faster convergence of our COHESION method is obviously observed, which suggests the advantage of COHESION in training efficiency while maintaining superior recommendation accuracy. This enhanced performance can be attributed to our innovative dual-stage fusion strategy, which effectively mitigates the influence of irrelevant information across modalities and brings a positive effect for model optimization to achieve fast convergence. Our dual-stage fusion strategy not only expedites the convergence speed but also enhances recommendation performance.

\subsection{Hyper-parameter Sensitivity Study (RQ6)}

\subsubsection{Number of Layer $L$ of Heterogeneous Graph}
\begin{figure}[h]
\vskip -0.1in
    \centering
    \subfigure[Baby] {
        \includegraphics[width=0.32\linewidth]{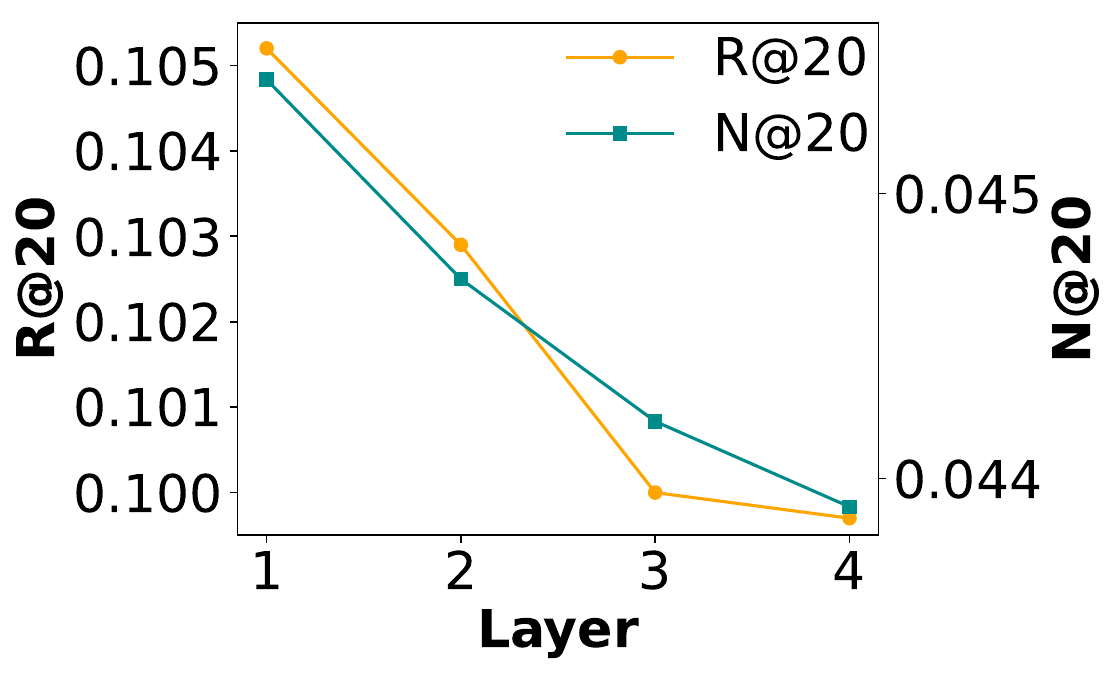}
        \label{fig.BabyLayer}
        }
    \hspace{-4.4mm}
    \subfigure[Sports] {
        \includegraphics[width=0.32\linewidth]{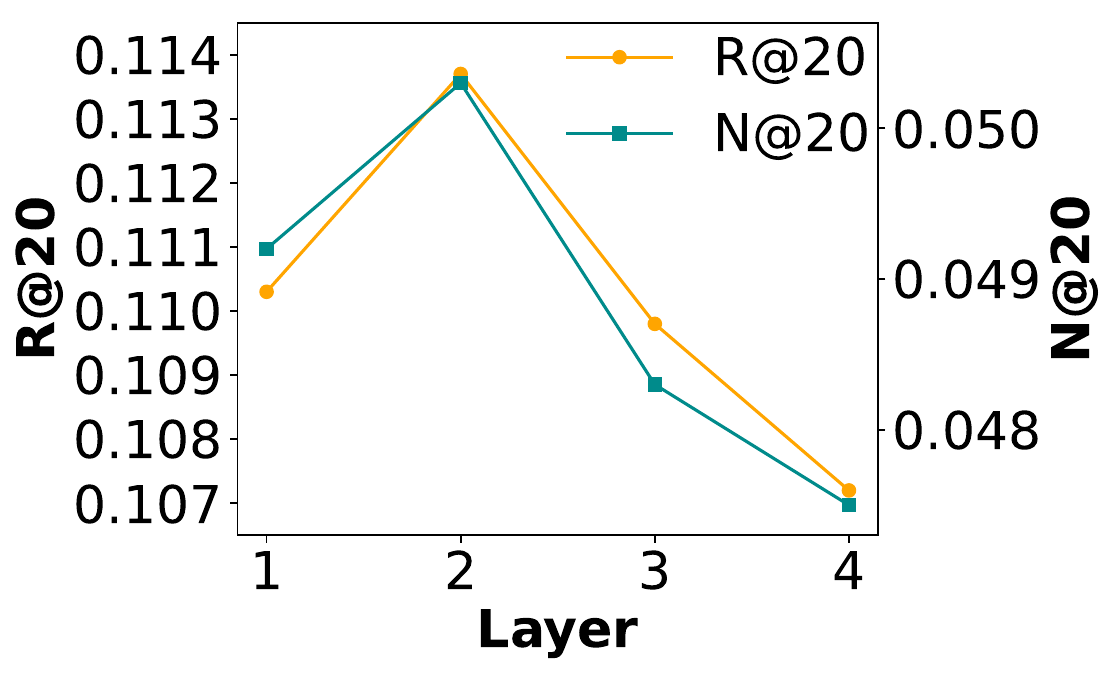}
        \label{fig.SportsLayer}
        }
    \hspace{-4.4mm}
    \subfigure[Clothing] {
        \includegraphics[width=0.32\linewidth]{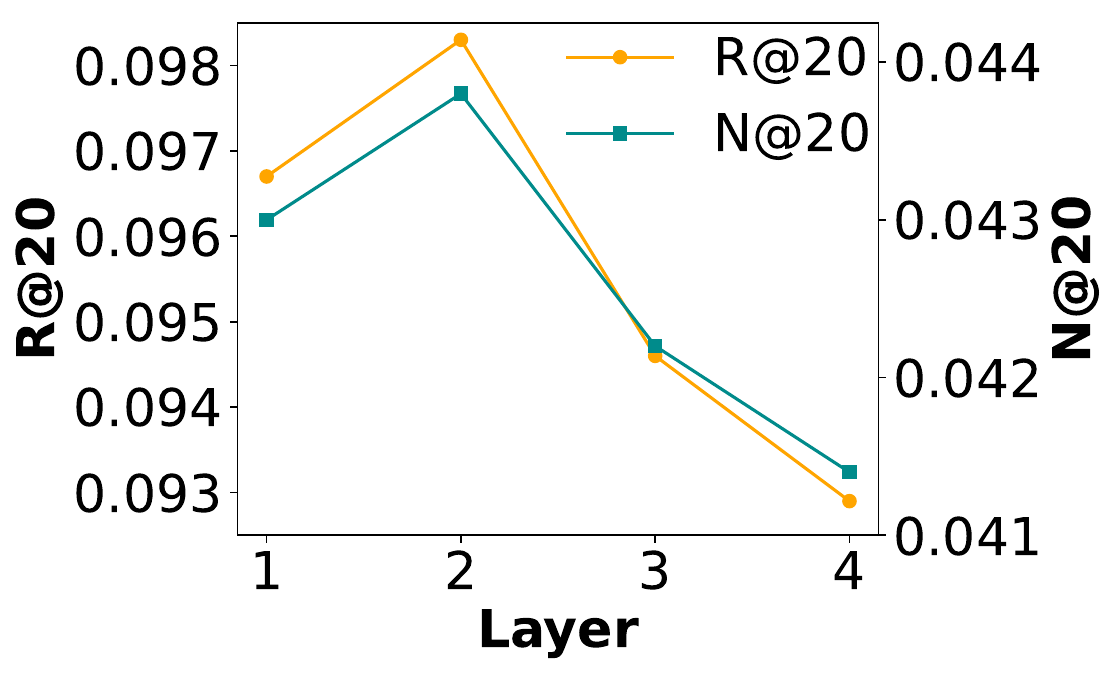}
        \label{fig.ClothingLayer}
        }
        \vskip -0.2in
    \caption{Effect of the number of heterogeneous graph layers.}
    \label{Layers}     
\end{figure}

As shown in Fig.~\ref{Layers}, we find that the optimal number of layer $L$ for Baby, Sports, and Clothing datasets are 1, 2, and 2, respectively. The reason is that the degree of association between users and items is smaller in the Baby dataset than in the Sports and Clothing datasets. It is important to emphasize that for all other multimodal recommendation methods using GCN, the best number of layers of GCN is 1 for all these three datasets. It verifies that residual-based GCN can alleviate the over-smoothing problem to some extent.

\begin{figure}[!h]
    \centering
    \vskip -0.25in
    \subfigure[Baby] {
        \label{fig:babyHeatmap}
        \includegraphics[width=0.32\linewidth]{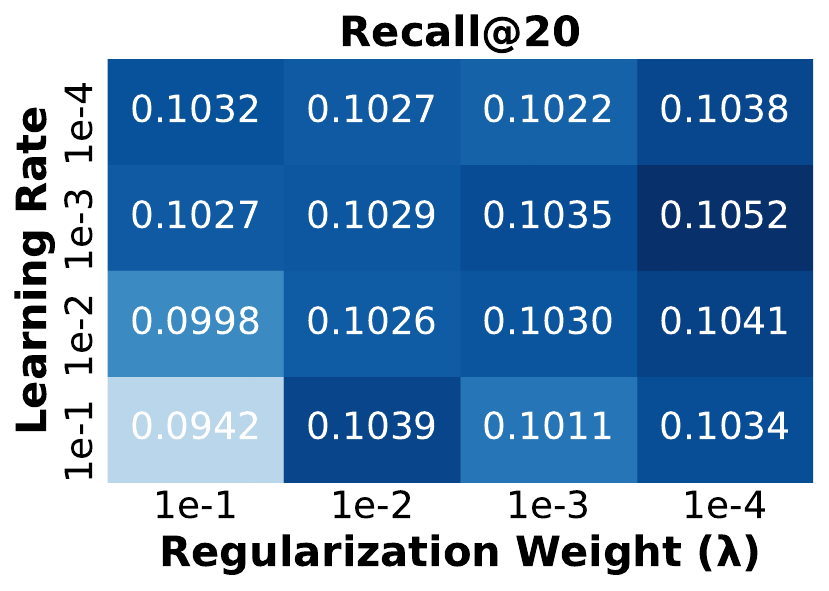}
        } 
    \hspace{-4.4mm}
    \subfigure[Sports] {
        \label{fig:sportsHeatmap}
        \includegraphics[width=0.32\linewidth]{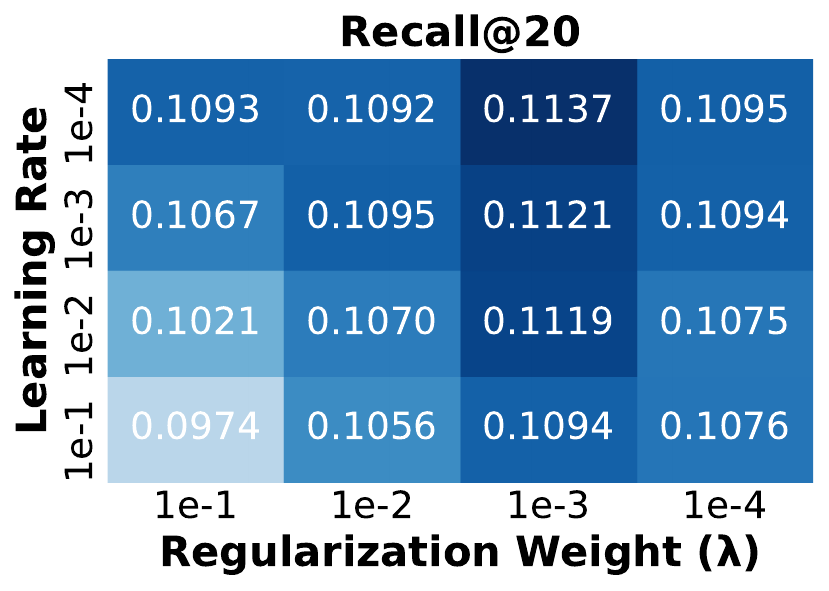}
        }   
    \hspace{-4.4mm}
    \subfigure[Clothing] {
        \label{fig:clothingHeatmap}
        \includegraphics[width=0.32\linewidth]{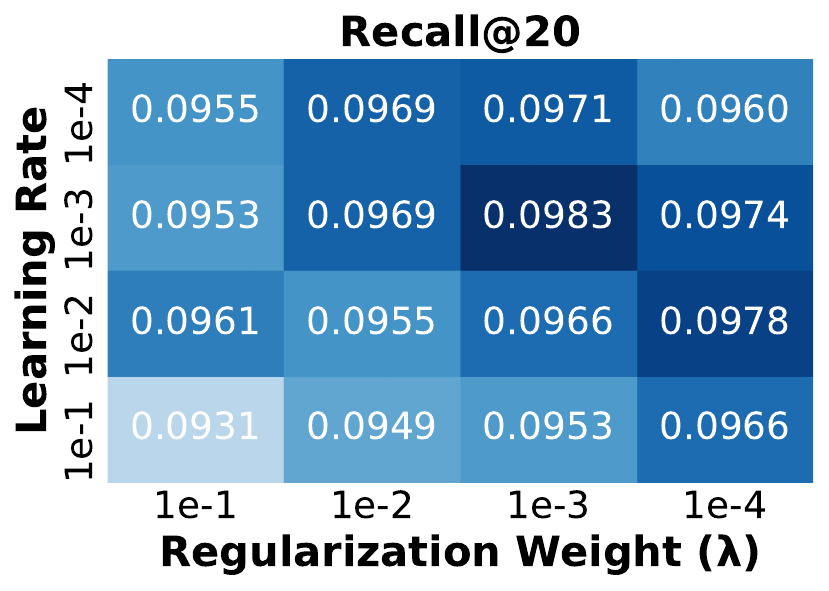}
        }   
        \vskip -0.2in
    \caption{Performance of COHESION with respect to different regularization weight $\lambda$ and learning rate.}   
    \label{fig:5}  
     \vskip -0.1in
\end{figure}

\subsubsection{The Learning Rate and Regularization Weight}
Fig.~\ref{fig:5} shows the results of COHESION under different learning rates and regularization weights on Baby, Sports, and Clothing datasets. We can observe that 1e-3, 1e-4, and 1e-3 are the suggested learning rates for the Baby, Sports, and Clothing datasets, respectively. Moreover, the suggested regularization weights are 1e-4, 1e-3, and 1e-3 for the Baby, Sports, and Clothing datasets, respectively.
\section{Conclusion}
In this paper, we examine the limitations of the fusion strategy in previous works and further reveal the complementary capability between fusion strategy and representation learning. To this end, we propose a composite graph convolutional network with a dual-stage fusion strategy for multimodal recommendation, named COHESION. More specifically, our well-designed composite graph convolutional network achieves outstanding representation learning ability by leveraging both heterogeneous and homogeneous graphs. Moreover, our tailored dual-stage fusion strategy further liberates the representation learning ability of our composite graph convolutional network by mitigating the negative effect of irrelevant information among modalities. Finally, we propose an adaptive optimization to control the balance of attention among all modalities. We conduct comprehensive experiments and ablation studies to demonstrate the effectiveness of our COHESION.



\balance

\begin{thebibliography}{42}


\ifx \showCODEN    \undefined \def \showCODEN     #1{\unskip}     \fi
\ifx \showDOI      \undefined \def \showDOI       #1{#1}\fi
\ifx \showISBNx    \undefined \def \showISBNx     #1{\unskip}     \fi
\ifx \showISBNxiii \undefined \def \showISBNxiii  #1{\unskip}     \fi
\ifx \showISSN     \undefined \def \showISSN      #1{\unskip}     \fi
\ifx \showLCCN     \undefined \def \showLCCN      #1{\unskip}     \fi
\ifx \shownote     \undefined \def \shownote      #1{#1}          \fi
\ifx \showarticletitle \undefined \def \showarticletitle #1{#1}   \fi
\ifx \showURL      \undefined \def \showURL       {\relax}        \fi
\providecommand\bibfield[2]{#2}
\providecommand\bibinfo[2]{#2}
\providecommand\natexlab[1]{#1}
\providecommand\showeprint[2][]{arXiv:#2}

\bibitem[Bertschinger et~al\mbox{.}(2014)]%
        {bertschinger2014quantifying}
\bibfield{author}{\bibinfo{person}{Nils Bertschinger}, \bibinfo{person}{Johannes Rauh}, \bibinfo{person}{Eckehard Olbrich}, \bibinfo{person}{J{\"u}rgen Jost}, {and} \bibinfo{person}{Nihat Ay}.} \bibinfo{year}{2014}\natexlab{}.
\newblock \showarticletitle{Quantifying unique information}.
\newblock \bibinfo{journal}{\emph{Entropy}} \bibinfo{volume}{16}, \bibinfo{number}{4} (\bibinfo{year}{2014}), \bibinfo{pages}{2161--2183}.
\newblock


\bibitem[Chen et~al\mbox{.}(2017)]%
        {chen2017attentive}
\bibfield{author}{\bibinfo{person}{Jingyuan Chen}, \bibinfo{person}{Hanwang Zhang}, \bibinfo{person}{Xiangnan He}, \bibinfo{person}{Liqiang Nie}, \bibinfo{person}{Wei Liu}, {and} \bibinfo{person}{Tat-Seng Chua}.} \bibinfo{year}{2017}\natexlab{}.
\newblock \showarticletitle{Attentive collaborative filtering: Multimedia recommendation with item-and component-level attention}. In \bibinfo{booktitle}{\emph{Proceedings of the 40th International ACM SIGIR conference on Research and Development in Information Retrieval}}. \bibinfo{pages}{335--344}.
\newblock


\bibitem[Chen et~al\mbox{.}(2025)]%
        {chen2025don}
\bibfield{author}{\bibinfo{person}{Zheyu Chen}, \bibinfo{person}{Jinfeng Xu}, {and} \bibinfo{person}{Haibo Hu}.} \bibinfo{year}{2025}\natexlab{}.
\newblock \showarticletitle{Don’t Lose Yourself: Boosting Multimodal Recommendation via Reducing Node-neighbor Discrepancy in Graph Convolutional Network}. In \bibinfo{booktitle}{\emph{ICASSP 2025-2025 IEEE International Conference on Acoustics, Speech and Signal Processing (ICASSP)}}. IEEE, \bibinfo{pages}{1--5}.
\newblock


\bibitem[Dong et~al\mbox{.}(2024)]%
        {dong2024prompt}
\bibfield{author}{\bibinfo{person}{Xue Dong}, \bibinfo{person}{Xuemeng Song}, \bibinfo{person}{Minghui Tian}, {and} \bibinfo{person}{Linmei Hu}.} \bibinfo{year}{2024}\natexlab{}.
\newblock \showarticletitle{Prompt-based and weak-modality enhanced multimodal recommendation}.
\newblock \bibinfo{journal}{\emph{Information Fusion}}  \bibinfo{volume}{101} (\bibinfo{year}{2024}), \bibinfo{pages}{101989}.
\newblock


\bibitem[Dufumier et~al\mbox{.}(2024)]%
        {dufumier2024align}
\bibfield{author}{\bibinfo{person}{Benoit Dufumier}, \bibinfo{person}{Javiera Castillo-Navarro}, \bibinfo{person}{Devis Tuia}, {and} \bibinfo{person}{Jean-Philippe Thiran}.} \bibinfo{year}{2024}\natexlab{}.
\newblock \showarticletitle{What to align in multimodal contrastive learning?}
\newblock \bibinfo{journal}{\emph{arXiv preprint arXiv:2409.07402}} (\bibinfo{year}{2024}).
\newblock


\bibitem[Glorot and Bengio(2010)]%
        {glorot2010understanding}
\bibfield{author}{\bibinfo{person}{Xavier Glorot} {and} \bibinfo{person}{Yoshua Bengio}.} \bibinfo{year}{2010}\natexlab{}.
\newblock \showarticletitle{Understanding the difficulty of training deep feedforward neural networks}. In \bibinfo{booktitle}{\emph{Proceedings of the thirteenth international conference on artificial intelligence and statistics}}. JMLR Workshop and Conference Proceedings, \bibinfo{pages}{249--256}.
\newblock


\bibitem[Guo et~al\mbox{.}(2024)]%
        {guo2024lgmrec}
\bibfield{author}{\bibinfo{person}{Zhiqiang Guo}, \bibinfo{person}{Jianjun Li}, \bibinfo{person}{Guohui Li}, \bibinfo{person}{Chaoyang Wang}, \bibinfo{person}{Si Shi}, {and} \bibinfo{person}{Bin Ruan}.} \bibinfo{year}{2024}\natexlab{}.
\newblock \showarticletitle{LGMRec: Local and Global Graph Learning for Multimodal Recommendation}. In \bibinfo{booktitle}{\emph{Proceedings of the AAAI Conference on Artificial Intelligence}}, Vol.~\bibinfo{volume}{38}. \bibinfo{pages}{8454--8462}.
\newblock


\bibitem[He and McAuley(2016)]%
        {he2016vbpr}
\bibfield{author}{\bibinfo{person}{Ruining He} {and} \bibinfo{person}{Julian McAuley}.} \bibinfo{year}{2016}\natexlab{}.
\newblock \showarticletitle{VBPR: visual bayesian personalized ranking from implicit feedback}. In \bibinfo{booktitle}{\emph{Proceedings of the AAAI conference on artificial intelligence}}, Vol.~\bibinfo{volume}{30}.
\newblock


\bibitem[He et~al\mbox{.}(2020)]%
        {he2020lightgcn}
\bibfield{author}{\bibinfo{person}{Xiangnan He}, \bibinfo{person}{Kuan Deng}, \bibinfo{person}{Xiang Wang}, \bibinfo{person}{Yan Li}, \bibinfo{person}{Yongdong Zhang}, {and} \bibinfo{person}{Meng Wang}.} \bibinfo{year}{2020}\natexlab{}.
\newblock \showarticletitle{Lightgcn: Simplifying and powering graph convolution network for recommendation}. In \bibinfo{booktitle}{\emph{Proceedings of the 43rd International ACM SIGIR conference on research and development in Information Retrieval}}. \bibinfo{pages}{639--648}.
\newblock


\bibitem[Jiang et~al\mbox{.}(2024)]%
        {jiang2024diffmm}
\bibfield{author}{\bibinfo{person}{Yangqin Jiang}, \bibinfo{person}{Lianghao Xia}, \bibinfo{person}{Wei Wei}, \bibinfo{person}{Da Luo}, \bibinfo{person}{Kangyi Lin}, {and} \bibinfo{person}{Chao Huang}.} \bibinfo{year}{2024}\natexlab{}.
\newblock \showarticletitle{DiffMM: Multi-Modal Diffusion Model for Recommendation}.
\newblock  (\bibinfo{year}{2024}).
\newblock


\bibitem[Kim et~al\mbox{.}(2022)]%
        {kim2022mario}
\bibfield{author}{\bibinfo{person}{Taeri Kim}, \bibinfo{person}{Yeon-Chang Lee}, \bibinfo{person}{Kijung Shin}, {and} \bibinfo{person}{Sang-Wook Kim}.} \bibinfo{year}{2022}\natexlab{}.
\newblock \showarticletitle{MARIO: modality-aware attention and modality-preserving decoders for multimedia recommendation}. In \bibinfo{booktitle}{\emph{Proceedings of the 31st ACM International Conference on Information \& Knowledge Management}}. \bibinfo{pages}{993--1002}.
\newblock


\bibitem[Kim et~al\mbox{.}(2024)]%
        {kim2024monet}
\bibfield{author}{\bibinfo{person}{Yungi Kim}, \bibinfo{person}{Taeri Kim}, \bibinfo{person}{Won-Yong Shin}, {and} \bibinfo{person}{Sang-Wook Kim}.} \bibinfo{year}{2024}\natexlab{}.
\newblock \showarticletitle{MONET: Modality-Embracing Graph Convolutional Network and Target-Aware Attention for Multimedia Recommendation}. In \bibinfo{booktitle}{\emph{Proceedings of the 17th ACM International Conference on Web Search and Data Mining}}. \bibinfo{pages}{332--340}.
\newblock


\bibitem[Kingma and Ba(2014)]%
        {kingma2014adam}
\bibfield{author}{\bibinfo{person}{Diederik~P Kingma} {and} \bibinfo{person}{Jimmy Ba}.} \bibinfo{year}{2014}\natexlab{}.
\newblock \showarticletitle{Adam: A method for stochastic optimization}.
\newblock \bibinfo{journal}{\emph{arXiv preprint arXiv:1412.6980}} (\bibinfo{year}{2014}).
\newblock


\bibitem[Koren et~al\mbox{.}(2021)]%
        {koren2021advances}
\bibfield{author}{\bibinfo{person}{Yehuda Koren}, \bibinfo{person}{Steffen Rendle}, {and} \bibinfo{person}{Robert Bell}.} \bibinfo{year}{2021}\natexlab{}.
\newblock \showarticletitle{Advances in collaborative filtering}.
\newblock \bibinfo{journal}{\emph{Recommender systems handbook}} (\bibinfo{year}{2021}), \bibinfo{pages}{91--142}.
\newblock


\bibitem[Lei et~al\mbox{.}(2023)]%
        {lei2023learning}
\bibfield{author}{\bibinfo{person}{Fei Lei}, \bibinfo{person}{Zhongqi Cao}, \bibinfo{person}{Yuning Yang}, \bibinfo{person}{Yibo Ding}, {and} \bibinfo{person}{Cong Zhang}.} \bibinfo{year}{2023}\natexlab{}.
\newblock \showarticletitle{Learning the User’s Deeper Preferences for Multi-modal Recommendation Systems}.
\newblock \bibinfo{journal}{\emph{ACM Transactions on Multimedia Computing, Communications and Applications}} \bibinfo{volume}{19}, \bibinfo{number}{3s} (\bibinfo{year}{2023}), \bibinfo{pages}{1--18}.
\newblock


\bibitem[McAuley et~al\mbox{.}(2015)]%
        {mcauley2015image}
\bibfield{author}{\bibinfo{person}{Julian McAuley}, \bibinfo{person}{Christopher Targett}, \bibinfo{person}{Qinfeng Shi}, {and} \bibinfo{person}{Anton Van Den~Hengel}.} \bibinfo{year}{2015}\natexlab{}.
\newblock \showarticletitle{Image-based recommendations on styles and substitutes}. In \bibinfo{booktitle}{\emph{Proceedings of the 38th international ACM SIGIR conference on research and development in information retrieval}}. \bibinfo{pages}{43--52}.
\newblock


\bibitem[Mu et~al\mbox{.}(2022)]%
        {mu2022learning}
\bibfield{author}{\bibinfo{person}{Zongshen Mu}, \bibinfo{person}{Yueting Zhuang}, \bibinfo{person}{Jie Tan}, \bibinfo{person}{Jun Xiao}, {and} \bibinfo{person}{Siliang Tang}.} \bibinfo{year}{2022}\natexlab{}.
\newblock \showarticletitle{Learning hybrid behavior patterns for multimedia recommendation}. In \bibinfo{booktitle}{\emph{Proceedings of the 30th ACM International Conference on Multimedia}}. \bibinfo{pages}{376--384}.
\newblock


\bibitem[Rendle et~al\mbox{.}(2012)]%
        {rendle2012bpr}
\bibfield{author}{\bibinfo{person}{Steffen Rendle}, \bibinfo{person}{Christoph Freudenthaler}, \bibinfo{person}{Zeno Gantner}, {and} \bibinfo{person}{Lars Schmidt-Thieme}.} \bibinfo{year}{2012}\natexlab{}.
\newblock \showarticletitle{BPR: Bayesian personalized ranking from implicit feedback}.
\newblock \bibinfo{journal}{\emph{arXiv preprint arXiv:1205.2618}} (\bibinfo{year}{2012}).
\newblock


\bibitem[Tang et~al\mbox{.}(2019)]%
        {tang2019adversarial}
\bibfield{author}{\bibinfo{person}{Jinhui Tang}, \bibinfo{person}{Xiaoyu Du}, \bibinfo{person}{Xiangnan He}, \bibinfo{person}{Fajie Yuan}, \bibinfo{person}{Qi Tian}, {and} \bibinfo{person}{Tat-Seng Chua}.} \bibinfo{year}{2019}\natexlab{}.
\newblock \showarticletitle{Adversarial training towards robust multimedia recommender system}.
\newblock \bibinfo{journal}{\emph{IEEE Transactions on Knowledge and Data Engineering}} \bibinfo{volume}{32}, \bibinfo{number}{5} (\bibinfo{year}{2019}), \bibinfo{pages}{855--867}.
\newblock


\bibitem[Tao et~al\mbox{.}(2022)]%
        {tao2022self}
\bibfield{author}{\bibinfo{person}{Zhulin Tao}, \bibinfo{person}{Xiaohao Liu}, \bibinfo{person}{Yewei Xia}, \bibinfo{person}{Xiang Wang}, \bibinfo{person}{Lifang Yang}, \bibinfo{person}{Xianglin Huang}, {and} \bibinfo{person}{Tat-Seng Chua}.} \bibinfo{year}{2022}\natexlab{}.
\newblock \showarticletitle{Self-supervised learning for multimedia recommendation}.
\newblock \bibinfo{journal}{\emph{IEEE Transactions on Multimedia}} (\bibinfo{year}{2022}).
\newblock


\bibitem[Tran and Lauw(2022)]%
        {tran2022aligning}
\bibfield{author}{\bibinfo{person}{Nhu-Thuat Tran} {and} \bibinfo{person}{Hady~W Lauw}.} \bibinfo{year}{2022}\natexlab{}.
\newblock \showarticletitle{Aligning Dual Disentangled User Representations from Ratings and Textual Content}. In \bibinfo{booktitle}{\emph{Proceedings of the 28th ACM SIGKDD Conference on Knowledge Discovery and Data Mining}}. \bibinfo{pages}{1798--1806}.
\newblock


\bibitem[Van~der Maaten and Hinton(2008)]%
        {van2008visualizing}
\bibfield{author}{\bibinfo{person}{Laurens Van~der Maaten} {and} \bibinfo{person}{Geoffrey Hinton}.} \bibinfo{year}{2008}\natexlab{}.
\newblock \showarticletitle{Visualizing data using t-SNE.}
\newblock \bibinfo{journal}{\emph{Journal of machine learning research}} \bibinfo{volume}{9}, \bibinfo{number}{11} (\bibinfo{year}{2008}).
\newblock


\bibitem[Wang et~al\mbox{.}(2021)]%
        {wang2021dualgnn}
\bibfield{author}{\bibinfo{person}{Qifan Wang}, \bibinfo{person}{Yinwei Wei}, \bibinfo{person}{Jianhua Yin}, \bibinfo{person}{Jianlong Wu}, \bibinfo{person}{Xuemeng Song}, {and} \bibinfo{person}{Liqiang Nie}.} \bibinfo{year}{2021}\natexlab{}.
\newblock \showarticletitle{Dualgnn: Dual graph neural network for multimedia recommendation}.
\newblock \bibinfo{journal}{\emph{IEEE Transactions on Multimedia}} (\bibinfo{year}{2021}).
\newblock


\bibitem[Wei et~al\mbox{.}(2020)]%
        {wei2020graph}
\bibfield{author}{\bibinfo{person}{Yinwei Wei}, \bibinfo{person}{Xiang Wang}, \bibinfo{person}{Liqiang Nie}, \bibinfo{person}{Xiangnan He}, {and} \bibinfo{person}{Tat-Seng Chua}.} \bibinfo{year}{2020}\natexlab{}.
\newblock \showarticletitle{Graph-refined convolutional network for multimedia recommendation with implicit feedback}. In \bibinfo{booktitle}{\emph{Proceedings of the 28th ACM international conference on multimedia}}. \bibinfo{pages}{3541--3549}.
\newblock


\bibitem[Wei et~al\mbox{.}(2019)]%
        {wei2019mmgcn}
\bibfield{author}{\bibinfo{person}{Yinwei Wei}, \bibinfo{person}{Xiang Wang}, \bibinfo{person}{Liqiang Nie}, \bibinfo{person}{Xiangnan He}, \bibinfo{person}{Richang Hong}, {and} \bibinfo{person}{Tat-Seng Chua}.} \bibinfo{year}{2019}\natexlab{}.
\newblock \showarticletitle{MMGCN: Multi-modal graph convolution network for personalized recommendation of micro-video}. In \bibinfo{booktitle}{\emph{Proceedings of the 27th ACM international conference on multimedia}}. \bibinfo{pages}{1437--1445}.
\newblock


\bibitem[Williams and Beer(2010)]%
        {williams2010nonnegative}
\bibfield{author}{\bibinfo{person}{Paul~L Williams} {and} \bibinfo{person}{Randall~D Beer}.} \bibinfo{year}{2010}\natexlab{}.
\newblock \showarticletitle{Nonnegative decomposition of multivariate information}.
\newblock \bibinfo{journal}{\emph{arXiv preprint arXiv:1004.2515}} (\bibinfo{year}{2010}).
\newblock


\bibitem[Xu et~al\mbox{.}(2024a)]%
        {xu2024aligngroup}
\bibfield{author}{\bibinfo{person}{Jinfeng Xu}, \bibinfo{person}{Zheyu Chen}, \bibinfo{person}{Jinze Li}, \bibinfo{person}{Shuo Yang}, \bibinfo{person}{Hewei Wang}, {and} \bibinfo{person}{Edith~CH Ngai}.} \bibinfo{year}{2024}\natexlab{a}.
\newblock \showarticletitle{AlignGroup: Learning and Aligning Group Consensus with Member Preferences for Group Recommendation}. In \bibinfo{booktitle}{\emph{Proceedings of the 33rd ACM International Conference on Information and Knowledge Management}}. \bibinfo{pages}{2682--2691}.
\newblock


\bibitem[Xu et~al\mbox{.}(2024b)]%
        {xu2024fourierkan}
\bibfield{author}{\bibinfo{person}{Jinfeng Xu}, \bibinfo{person}{Zheyu Chen}, \bibinfo{person}{Jinze Li}, \bibinfo{person}{Shuo Yang}, \bibinfo{person}{Wei Wang}, \bibinfo{person}{Xiping Hu}, {and} \bibinfo{person}{Edith C-H Ngai}.} \bibinfo{year}{2024}\natexlab{b}.
\newblock \showarticletitle{FourierKAN-GCF: Fourier Kolmogorov-Arnold Network--An Effective and Efficient Feature Transformation for Graph Collaborative Filtering}.
\newblock \bibinfo{journal}{\emph{arXiv preprint arXiv:2406.01034}} (\bibinfo{year}{2024}).
\newblock


\bibitem[Xu et~al\mbox{.}(2024c)]%
        {xu2024improving}
\bibfield{author}{\bibinfo{person}{Jinfeng Xu}, \bibinfo{person}{Zheyu Chen}, \bibinfo{person}{Zixiao Ma}, \bibinfo{person}{Jiyi Liu}, {and} \bibinfo{person}{Edith~CH Ngai}.} \bibinfo{year}{2024}\natexlab{c}.
\newblock \showarticletitle{Improving Consumer Experience With Pre-Purify Temporal-Decay Memory-Based Collaborative Filtering Recommendation for Graduate School Application}.
\newblock \bibinfo{journal}{\emph{IEEE Transactions on Consumer Electronics}} (\bibinfo{year}{2024}).
\newblock


\bibitem[Xu et~al\mbox{.}(2024d)]%
        {xu2024mentor}
\bibfield{author}{\bibinfo{person}{Jinfeng Xu}, \bibinfo{person}{Zheyu Chen}, \bibinfo{person}{Shuo Yang}, \bibinfo{person}{Jinze Li}, \bibinfo{person}{Hewei Wang}, {and} \bibinfo{person}{Edith C-H Ngai}.} \bibinfo{year}{2024}\natexlab{d}.
\newblock \showarticletitle{MENTOR: Multi-level Self-supervised Learning for Multimodal Recommendation}.
\newblock \bibinfo{journal}{\emph{arXiv preprint arXiv:2402.19407}} (\bibinfo{year}{2024}).
\newblock


\bibitem[Xu et~al\mbox{.}(2025)]%
        {xu2025survey}
\bibfield{author}{\bibinfo{person}{Jinfeng Xu}, \bibinfo{person}{Zheyu Chen}, \bibinfo{person}{Shuo Yang}, \bibinfo{person}{Jinze Li}, \bibinfo{person}{Wei Wang}, \bibinfo{person}{Xiping Hu}, \bibinfo{person}{Steven Hoi}, {and} \bibinfo{person}{Edith Ngai}.} \bibinfo{year}{2025}\natexlab{}.
\newblock \showarticletitle{A Survey on Multimodal Recommender Systems: Recent Advances and Future Directions}.
\newblock \bibinfo{journal}{\emph{arXiv preprint arXiv:2502.15711}} (\bibinfo{year}{2025}).
\newblock


\bibitem[Xu et~al\mbox{.}(2020)]%
        {xu2020reluplex}
\bibfield{author}{\bibinfo{person}{Jin Xu}, \bibinfo{person}{Zishan Li}, \bibinfo{person}{Bowen Du}, \bibinfo{person}{Miaomiao Zhang}, {and} \bibinfo{person}{Jing Liu}.} \bibinfo{year}{2020}\natexlab{}.
\newblock \showarticletitle{Reluplex made more practical: Leaky ReLU}. In \bibinfo{booktitle}{\emph{2020 IEEE Symposium on Computers and communications (ISCC)}}. IEEE, \bibinfo{pages}{1--7}.
\newblock


\bibitem[Xv et~al\mbox{.}(2024)]%
        {xv2024improving}
\bibfield{author}{\bibinfo{person}{Guipeng Xv}, \bibinfo{person}{Xinyu Li}, \bibinfo{person}{Ruobing Xie}, \bibinfo{person}{Chen Lin}, \bibinfo{person}{Chong Liu}, \bibinfo{person}{Feng Xia}, \bibinfo{person}{Zhanhui Kang}, {and} \bibinfo{person}{Leyu Lin}.} \bibinfo{year}{2024}\natexlab{}.
\newblock \showarticletitle{Improving Multi-modal Recommender Systems by Denoising and Aligning Multi-modal Content and User Feedback}. In \bibinfo{booktitle}{\emph{Proceedings of the 30th ACM SIGKDD Conference on Knowledge Discovery and Data Mining}}. \bibinfo{pages}{3645--3656}.
\newblock


\bibitem[Yi and Chen(2021)]%
        {yi2021multi}
\bibfield{author}{\bibinfo{person}{Jing Yi} {and} \bibinfo{person}{Zhenzhong Chen}.} \bibinfo{year}{2021}\natexlab{}.
\newblock \showarticletitle{Multi-modal variational graph auto-encoder for recommendation systems}.
\newblock \bibinfo{journal}{\emph{IEEE Transactions on Multimedia}}  \bibinfo{volume}{24} (\bibinfo{year}{2021}), \bibinfo{pages}{1067--1079}.
\newblock


\bibitem[Yu et~al\mbox{.}(2022)]%
        {yu2022graph}
\bibfield{author}{\bibinfo{person}{Junliang Yu}, \bibinfo{person}{Hongzhi Yin}, \bibinfo{person}{Xin Xia}, \bibinfo{person}{Tong Chen}, \bibinfo{person}{Lizhen Cui}, {and} \bibinfo{person}{Quoc Viet~Hung Nguyen}.} \bibinfo{year}{2022}\natexlab{}.
\newblock \showarticletitle{Are graph augmentations necessary? simple graph contrastive learning for recommendation}. In \bibinfo{booktitle}{\emph{Proceedings of the 45th international ACM SIGIR conference on research and development in information retrieval}}. \bibinfo{pages}{1294--1303}.
\newblock


\bibitem[Zhang et~al\mbox{.}(2021)]%
        {zhang2021mining}
\bibfield{author}{\bibinfo{person}{Jinghao Zhang}, \bibinfo{person}{Yanqiao Zhu}, \bibinfo{person}{Qiang Liu}, \bibinfo{person}{Shu Wu}, \bibinfo{person}{Shuhui Wang}, {and} \bibinfo{person}{Liang Wang}.} \bibinfo{year}{2021}\natexlab{}.
\newblock \showarticletitle{Mining latent structures for multimedia recommendation}. In \bibinfo{booktitle}{\emph{Proceedings of the 29th ACM International Conference on Multimedia}}. \bibinfo{pages}{3872--3880}.
\newblock


\bibitem[Zhou et~al\mbox{.}(2023c)]%
        {zhou2023comprehensive}
\bibfield{author}{\bibinfo{person}{Hongyu Zhou}, \bibinfo{person}{Xin Zhou}, \bibinfo{person}{Zhiwei Zeng}, \bibinfo{person}{Lingzi Zhang}, {and} \bibinfo{person}{Zhiqi Shen}.} \bibinfo{year}{2023}\natexlab{c}.
\newblock \showarticletitle{A comprehensive survey on multimodal recommender systems: Taxonomy, evaluation, and future directions}.
\newblock \bibinfo{journal}{\emph{arXiv preprint arXiv:2302.04473}} (\bibinfo{year}{2023}).
\newblock


\bibitem[Zhou et~al\mbox{.}(2023d)]%
        {zhou2023enhancing}
\bibfield{author}{\bibinfo{person}{Hongyu Zhou}, \bibinfo{person}{Xin Zhou}, \bibinfo{person}{Lingzi Zhang}, {and} \bibinfo{person}{Zhiqi Shen}.} \bibinfo{year}{2023}\natexlab{d}.
\newblock \showarticletitle{Enhancing dyadic relations with homogeneous graphs for multimodal recommendation}.
\newblock In \bibinfo{booktitle}{\emph{ECAI 2023}}. \bibinfo{publisher}{IOS Press}, \bibinfo{pages}{3123--3130}.
\newblock


\bibitem[Zhou(2023)]%
        {zhou2023mmrecsm}
\bibfield{author}{\bibinfo{person}{Xin Zhou}.} \bibinfo{year}{2023}\natexlab{}.
\newblock \showarticletitle{MMRec: Simplifying Multimodal Recommendation}.
\newblock \bibinfo{journal}{\emph{arXiv preprint arXiv:2302.03497}} (\bibinfo{year}{2023}).
\newblock


\bibitem[Zhou et~al\mbox{.}(2023a)]%
        {zhou2023layer}
\bibfield{author}{\bibinfo{person}{Xin Zhou}, \bibinfo{person}{Donghui Lin}, \bibinfo{person}{Yong Liu}, {and} \bibinfo{person}{Chunyan Miao}.} \bibinfo{year}{2023}\natexlab{a}.
\newblock \showarticletitle{Layer-refined graph convolutional networks for recommendation}. In \bibinfo{booktitle}{\emph{2023 IEEE 39th International Conference on Data Engineering (ICDE)}}. IEEE, \bibinfo{pages}{1247--1259}.
\newblock


\bibitem[Zhou and Shen(2023)]%
        {zhou2023tale}
\bibfield{author}{\bibinfo{person}{Xin Zhou} {and} \bibinfo{person}{Zhiqi Shen}.} \bibinfo{year}{2023}\natexlab{}.
\newblock \showarticletitle{A tale of two graphs: Freezing and denoising graph structures for multimodal recommendation}. In \bibinfo{booktitle}{\emph{Proceedings of the 31st ACM International Conference on Multimedia}}. \bibinfo{pages}{935--943}.
\newblock


\bibitem[Zhou et~al\mbox{.}(2023b)]%
        {zhou2023bootstrap}
\bibfield{author}{\bibinfo{person}{Xin Zhou}, \bibinfo{person}{Hongyu Zhou}, \bibinfo{person}{Yong Liu}, \bibinfo{person}{Zhiwei Zeng}, \bibinfo{person}{Chunyan Miao}, \bibinfo{person}{Pengwei Wang}, \bibinfo{person}{Yuan You}, {and} \bibinfo{person}{Feijun Jiang}.} \bibinfo{year}{2023}\natexlab{b}.
\newblock \showarticletitle{Bootstrap latent representations for multi-modal recommendation}. In \bibinfo{booktitle}{\emph{Proceedings of the ACM Web Conference 2023}}. \bibinfo{pages}{845--854}.
\newblock


\end{thebibliography}

\end{document}